\input{aipcheck.tex}

\def\selectedoptions{}
\ifx\empty\selectedoptions
  \def\selectedoptions{final}
\fi

\documentclass[
   \selectedoptions
  ]
  {aipproc}

\def\selectedlayoutstyle{6x9} 

\def\kpnnp{$K^+ \to \pi^+ \nu\bar\nu$}
\def\bkpnnp{$B(K^+ \to \pi^+ \nu\bar\nu)$}
\def\kpnn0{$K_L\to\pi^0\nu\bar\nu$}
\def\bkpnn0{$B(K_L\to\pi^0\nu\bar\nu)$}
\def\kmm{$K_L\to\mu^+\mu^-$}
\def\bkmm{$B(K_L\to\mu^+\mu^-)$}
\def\bkmmr{B(K_L\to\mu^+\mu^-)}
\def\kp0{$K_L \to \pi^0 \pi^0$}

\def\bkppr{B(K_L \to \pi^+ \pi^-)}
\def\kpll{$K_L \to \pi^0 \ell^+ \ell^-$}
\def\kpmm{$K_L \to \pi^0 \mu^+ \mu^-$}
\def\kpgg{$K_L \to \pi^0 \gamma\gamma$}

\def\be{\begin{equation}}
\def\ee{\end{equation}}
\def\bea{\begin{eqnarray}}
\def\eea{\end{eqnarray}}

\layoutstyle\selectedlayoutstyle

\SetInternalRegister\hbadness{8000} 

\newcommand\doingARLO[2][]{%
  \ifx\mmref\undefined #1\else #2\fi
}

\begin{document}

\title 
      []
      {Rare Kaon and Pion Decays\footnote{Lectures given at the PSI 
Summer School on Particle Physics 18-24 Aug 2002}
}


\author{Laurence Littenberg}{
  address={Brookhaven National Laboratory, Upton, NY 11973},
  email={littenbe@bnl.gov}
}

\copyrightyear  {2001}

\begin{abstract}
Recent results on rare kaon and pion decays are reviewed and prospects for 
future experiments are discussed.
\end{abstract}

\date{\today}

\maketitle

\section{Introduction}

The study of kaon and pion rare decays has three primary
motivations. The first is the search for physics beyond the Standard
Model (BSM).  Virtually all attempts to redress the theoretical
shortcomings of the Standard Model (SM) predict some degree of lepton flavor
violation (LFV).  Decays such as $K_L \to \mu^{\pm} e^{\mp}$ and
$K^+ \to \pi^+ \mu^+ e^-$ have excellent
experimental signatures and can consequently be pursued to remarkable
sensitivities.  These sensitivities correspond to extremely high energy scales
in models where the only suppression is that of the mass of the
exchanged field.  There are also theories that predict new particles
created in kaon or pion decay or the violation of symmetries other than lepton
flavor.

        The second motivation is the potential of decays that are allowed 
but that are extremely suppressed in the SM.  In several kaon decays,
the leading component
is a G.I.M.-suppressed\cite{Reference:GIM} one-loop process that 
is quite sensitive to fundamental SM parameters such as $V_{td}$.
These decays are also potentially very sensitive to BSM physics.
One interesting rare pion decay, $\pi^+ \to \pi^0 e^+ \nu_e$, that
is suppressed only by kinematics, can give a very clean measure of
$V_{ud}$, and possibly shed light on an apparent violation of
CKM unitarity.

           Finally there are long-distance-dominated
decays that can test theoretical techniques such as
chiral perturbation theory ($\chi$PT) that
purport to elucidate the low-energy behavior of
QCD.  Also, information from some of these decays is required to extract
fundamental information from certain of the one-loop processes.

        This field is quite active, as indicated by Table \ref{decays:a},
that lists the rare 
decays for which results have been forthcoming in the last couple of years,
as well as those that are under analysis.  In the face of such riches,
one must be quite selective in a short review such as this.

\begin{table}[h]
\begin{tabular}{llll} 
\hline
$K^+ \to \pi^+ \nu\bar\nu$ & $K_L \to \pi^0 \nu\bar\nu$ &
$K_L \to \pi^0 \mu^+\mu^-$ & $K_L \to \pi^0 e^+e^-$ \\
$K^+ \to \pi^+ \mu^+\mu^-$ & $K^+ \to \pi^+ e^+e^-$ &
$K_L \to  \mu^+\mu^-$ & $K_L \to  e^+e^-$ \\
$K^+ \to \pi^+ e^+ e^- \gamma$ & $K^+ \to \pi^+ \pi^0 \nu\bar\nu$ &
$K_L \to e^{\pm} e^{\mp} \mu^{\pm} \mu^{\mp}$ & $K^+ \to \pi^+ \pi^0 \gamma$ \\
$K_L \to \pi^+ \pi^- \gamma$ & $K_L \to \pi^+ \pi^- e^+ e^-$ &
$K^+ \to \pi^+ \pi^0 e^+ e^-$ & $K^+ \to \pi^0 \mu^+ \nu \gamma$ \\
$K_L \to \pi^0 \gamma \gamma$ & $K^+ \to \pi^+ \gamma \gamma$ &
$K^+ \to \mu^+ \nu \gamma$ & $K^+ \to e^+ \nu e^+ e^-$ \\
$K^+ \to \mu^+ \nu e^+ e^-$ & $K^+ \to e^+ \nu \mu^+\mu^-$ &
$K_L \to e^+ e^- \gamma$ & $K_L \to \mu^+ \mu^- \gamma$ \\
$K_L \to e^+ e^- \gamma\gamma$ & $K_L \to \mu^+ \mu^- \gamma\gamma$ &
$K_L \to e^+ e^- e^+ e^-$ & $K_L \to \pi^0 e^+ e^- \gamma$ \\
$K^+ \to \pi^+ \mu^+e^-$ & $K_L \to \pi^0 \mu^{\pm} e^{\mp}$ &
$K_L \to \mu^{\pm} e^{\mp}$ & $K^+ \to \pi^- \mu^+ e^+$ \\
$K^+ \to \pi^- e^+ e^+$ & $K^+ \to \pi^- \mu^+ \mu^+$ &
$K^+ \to \pi^+ X^0$ & $K_L \to e^{\pm} e^{\pm} \mu^{\mp} \mu^{\mp}$ \\
$K^+ \to \pi^+ \gamma$& $K_L \to \pi^0 \pi^0 e^+ e^-$ && \\
 $\pi^+ \to \pi^0 e^+ \nu_e$ & $\pi^+ \to e^+ \nu_e$ &
 $\pi^+ \to e^+ \nu_e \gamma$ & $\pi^+ \to e^+ \nu_e e^+ e^-$ \\
 $\pi^+ \to e^+ \nu_e \nu\bar\nu$ & $\pi^0 \to e^+ e^-$ &
 $\pi^0 \to e^+ e^- e^+ e^-$ &  $\pi^0 \to \nu\bar\nu$ \\ 
 $\pi^0 \to \gamma \nu\bar\nu$ &  $\pi^0 \to 3 \gamma$ &
 $\pi^0 \to \mu e$   &\\
\hline
\end{tabular}
\caption{\it Rare $K$ decay modes under recent or on-going study.}
\label{decays:a}
\end{table}

\section{Beyond the Standard Model}

\begin{figure}[h]
 \includegraphics[angle=0, height=.24\linewidth]{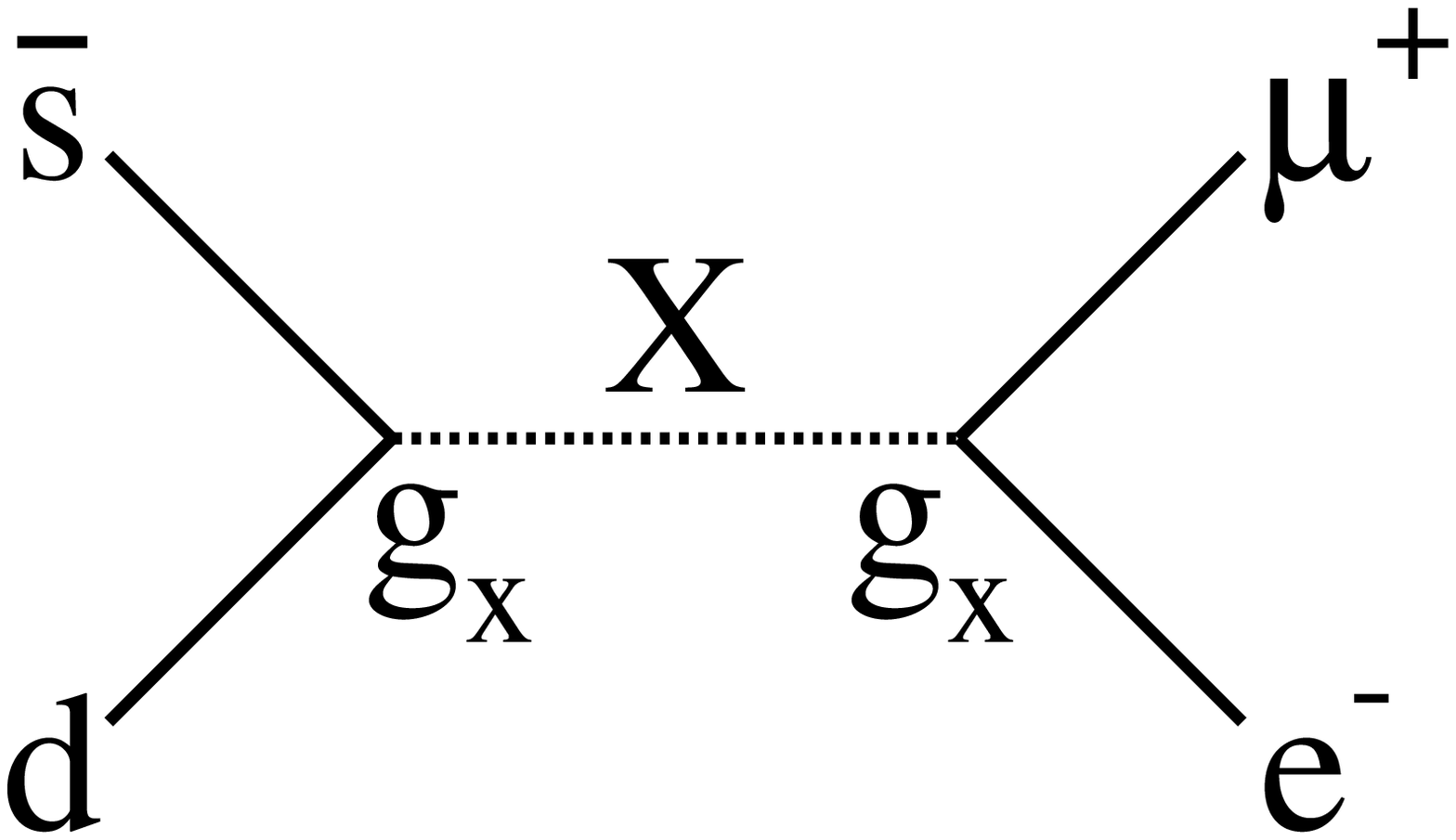}
 \includegraphics[angle=0, height=.24\linewidth]{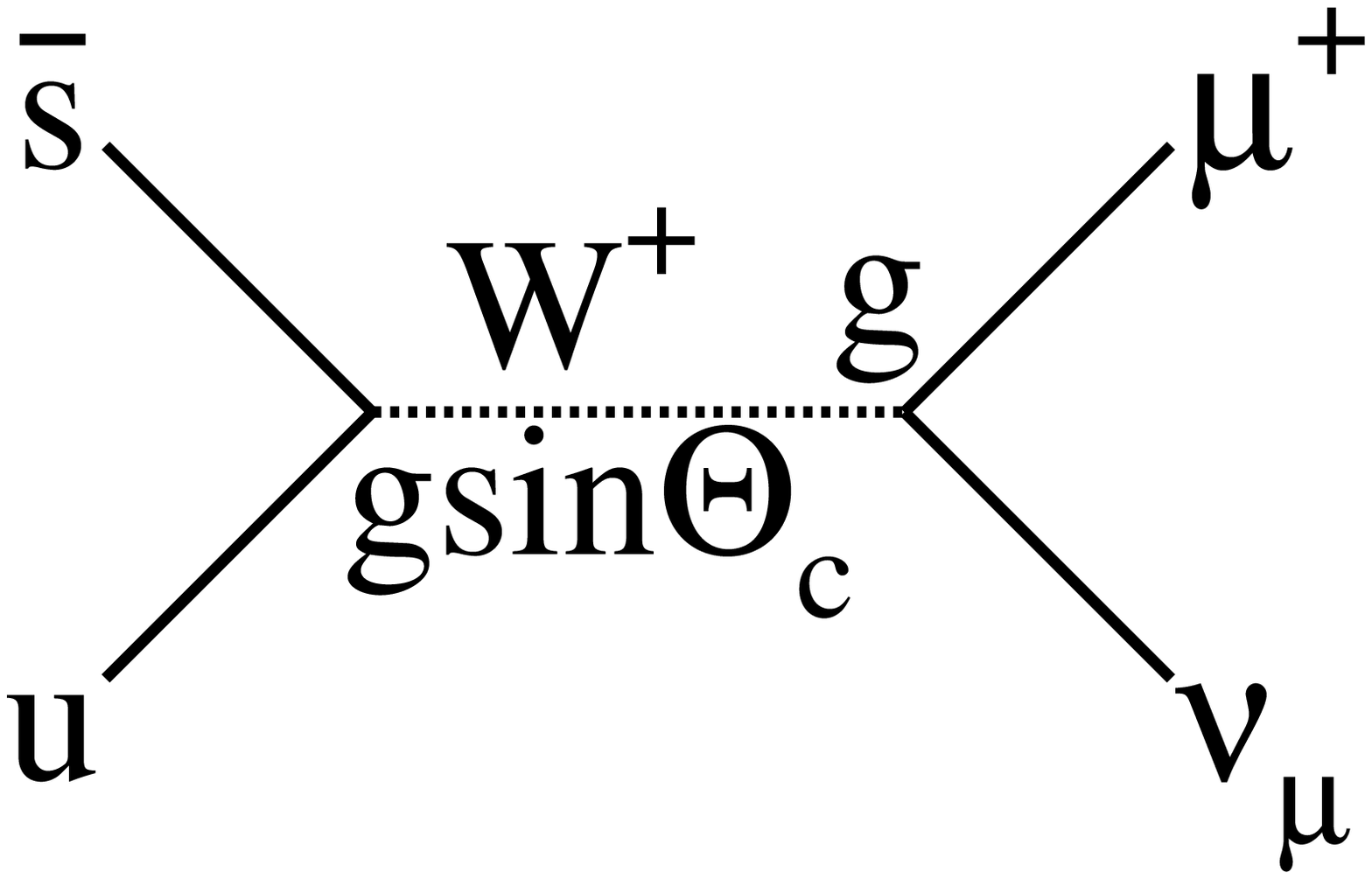}
  \caption{Horizontal gauge boson mediating $K_L \to \mu e$, compared with
$W$ mediating $K^+ \to \mu^+ \nu$.
    \label{fig:poster} }
\end{figure}

	The poster children for BSM probes in kaon decay are LFV
processes like $K_L \to \mu e$ and $K^+ \to \pi^+ \mu^+ e^-$.  In
principle, these can proceed through neutrino mixing, but the known
neutrino mixing parameters limit the rate through this mechanism to a
completely negligible level~\cite{Lee:1977ti}.  Thus the observation of
LFV in kaon decay would require a new mechanism.
Fig.\ref{fig:poster} shows $K_L \to \mu e$ mediated by a
hypothetical horizontal gauge boson $X$, compared with the
kinematically very similar process $K^+ \to \mu^+ \nu$ mediated by a
$W$ boson.

\noindent
Using measured values for $M_W$, the $K_L$ and $K^+$ decay rates and
$B(K^+ \to \mu^+ \nu)$, and assuming a $V-A$ form for the
new interaction, one can show~\cite{Cahn:1980kv}:
\be
M_X \approx 220\, TeV/c^2 \left[{g_X \over{g}}\right]^{1/4} \left[ {10^{-12} \over{B(K_L \to \mu e)}} \right]^{1/4}
\label{LFVcomp}
\ee so that truly formidable scales can be probed if $g_X \sim g$.  In addition
to this generic picture, there are specific models, such as extended
technicolor in which LFV at observable levels in kaon decays is 
quite natural~\cite{Eichten:1986eq}.

        There were a number of $K$ decay experiments primarily
dedicated to lepton flavor violation at the Brookhaven AGS during the
1990's.  These advanced the sensitivity to such processes by many
orders of magnitude.  In addition, several ``by-product'' results on
LFV and other BSM topics have emerged from the other kaon decay
experiments of this period.  Rare kaon decay experiments often also
yield results on $\pi^0$ decays, since these can readily be tagged,
{\it e.g.} via $K^+ \to \pi^+ \pi^0$ or $K_L \to \pi^+ \pi^- \pi^0$.
Table~\ref{BSM} summarizes the status of work on BSM probes in kaon
and pion decay.  The relative reach of these processes is best assessed
by comparing the partial rates rather than the branching ratios.  From
this table it is evident that pion decay is not yet a competitor to kaon decay
in probing LFV.

\begin{table}[h]
\begin{tabular}{|c|c|c|c|c|c|} \hline
Process & Violates & 90\% CL BR Limit & $\Gamma$ Limit (sec$^-1$) & Experiment & Reference \\
\hline
\hline
$K_L \to \mu e $ & LF& $4.7 \times 10^{-12}$ & $9.1 \times 10^{-5}$ &AGS-871 & \cite{Ambrose:1998us} \\
$K^+ \to \pi^+ \mu^+ e^-$ & LF &  $2.8 \times 10^{-11}$ & $2.3 \times 10^{-3}$ & AGS-865 &\cite{Appel:2000wg}  \\
$K^+ \to \pi^+ \mu^- e^+$ & LF, G &$5.2 \times 10^{-10}$ & $4.2 \times 10^{-2}$ & AGS-865 & \cite{Appel:2000tc} \\
$K_L \to \pi^0 \mu e$ & LF & $3.31 \times 10^{-10}$ & $6.4 \times 10^{-3}$ &  KTeV & \cite{Bellavance:2002}\\
$K^+ \to \pi^- e^+ e^+$ & LN, G &$6.4 \times 10^{-10}$ & $5.2 \times 10^{-2}$ & AGS-865 & \cite{Appel:2000tc} \\
$K^+ \to \pi^- \mu^+ \mu^+$ & LN, G &$3.0 \times 10^{-9}$ & $2.4 \times 10^{-1}$ & AGS-865 & \cite{Appel:2000tc} \\
$K^+ \to \pi^- \mu^+ e^+$ & LF, LN, G &$5.0 \times 10^{-10}$ & $4.0 \times 10^{-2}$ & AGS-865 & \cite{Appel:2000tc} \\
$K_L \to \mu^{\pm} \mu^{\pm} e^{\mp} e^{\mp} $ & LF, LN, G & $4.12 \times 10^{-11}$ & $8.0 \times 10^{-4}$ & KTeV & \cite{Hamm:2002vy}\\
$K^+ \to \pi^+ f^0$ & N & $5.9 \times 10^{-11}$ & $4.8 \times 10^{-3}$ & AGS-787 & \cite{Adler:2001xv}\\
$K^+ \to \pi^+ \gamma$ & H & $3.6 \times 10^{-7}$& $2.9 \times 10^1$ & AGS-787 & \cite{Adler:2001dt}\\
$\pi^0 \to \mu^+ e^-$ & LF & $3.8 \times 10^{-10}$ & $4.5 \times 10^6$ & AGS-865 & \cite{Appel:2000wg}\\
$\pi^0 \to \mu^- e^+$ & LF & $3.4 \times 10^{-9}$ & $4.0 \times 10^7$ & AGS-865 & \cite{Appel:2000tc}\\
$\pi^+ \to \mu^- e^+ e^+ \nu_e$ & LF & $1.6 \times 10^{-6}$ & $6.1 \times 10^1$ & JINR-SPEC & \cite
{Baranov:1991uj}\\
\hline
\end{tabular}
\caption{\it Current 90\% CL limits on  $K$ and $\pi$ decay modes violating the SM.  The violation
codes are ``LF'' for lepton flavor, ``LN'' for lepton number,
``G'' for generation number, \protect\cite{Cahn:1980kv}, ``H'' for helicity, ``N'' requires new particle}
\label{BSM}
\end{table}

It is clear from this table that any deviation from the SM must be
highly suppressed.  The kaon LFV probes in particular have become the
victims of their own success.  The specific theories they were
designed to probe have been killed or at least forced to retreat to
the point where meaningful tests in the kaon system would be very
difficult.  For example, although these decays provide the most
stringent limits on strangeness-changing R-violating couplings, the
minimal supersymmetric extension of the Standard Model predicts LFV in
kaon decay at levels far beyond the current experimental state of the
art \cite{Belyaev:2000xt}.  

\begin{figure}[t]
 \includegraphics[angle=0, height=.65\textheight]{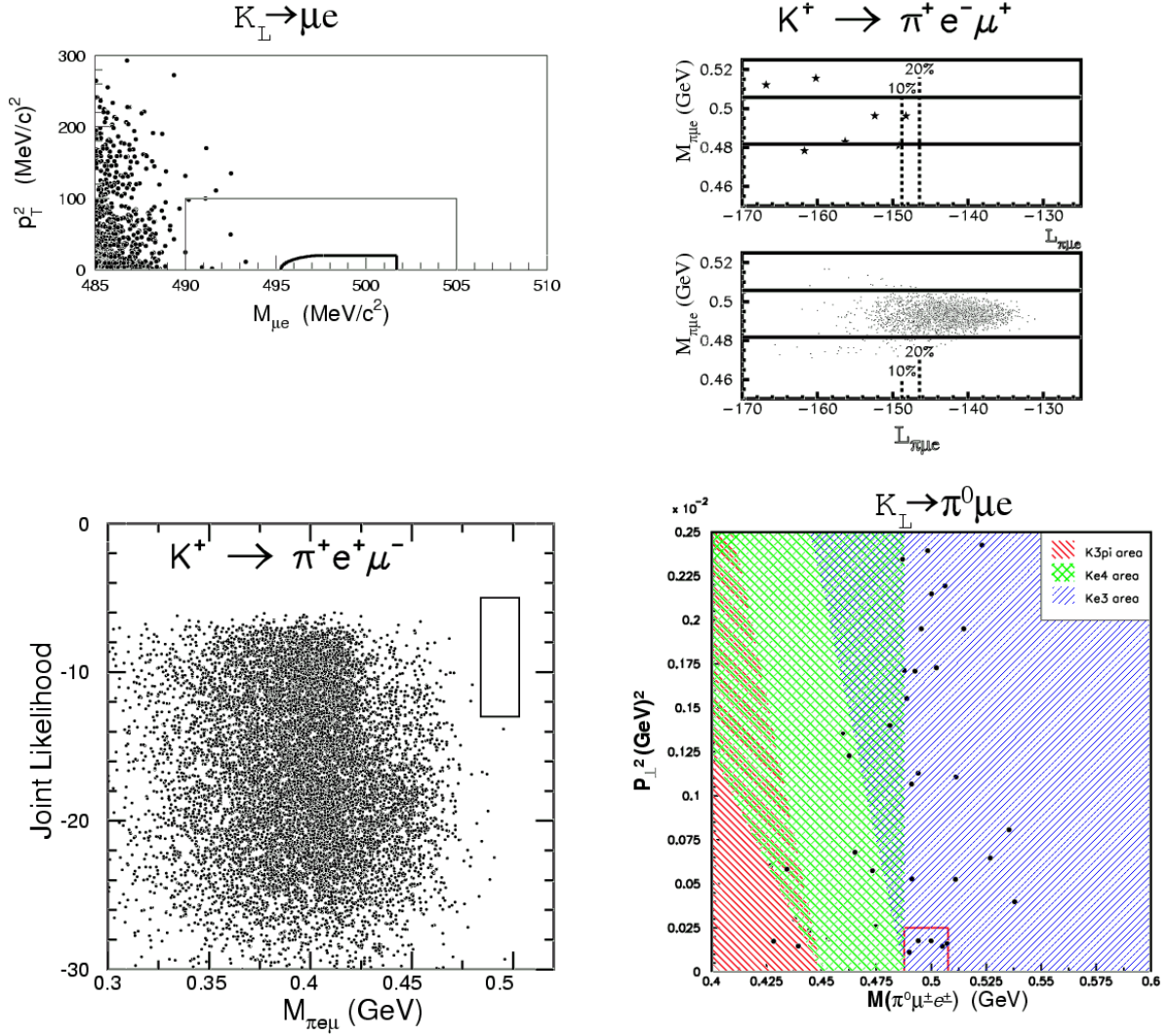}
  \caption{Signal planes showing candidates for LFV kaon decays from
recent experiments.
{\bf Top left:} $p_T^2$ vs $M_{\mu e}$ from Ref.~\protect\cite{Ambrose:1998us}, 
{\bf Top right:} $M_{\pi \mu e}$ vs Log likelihood from Ref.~\protect\cite{Appel:2000wg}
(lower plot shows signal Monte Carlo), 
{\bf Bottom left:} Joint likelihood vs $M_{\pi \mu e}$ from Ref.~\protect\cite{Appel:2000tc},
and {\bf Bottom right:} $P^2_{\perp}$ vs $M_{\pi \mu e}$ from Ref.~\protect\cite{Bellavance:2002}.
    \label{fig:boxes} }
\end{figure}

	Moreover both kaon flux and rejection of background are
becoming problematical.  Fig.\ref{fig:boxes} shows the signal planes
of four of the most sensitive LFV searches.  It is clear that
background either is already a problem or soon would be if the
sensitivity of these searches were increased.  Thus new techniques
will need to be developed to push such searches significantly
further.

Analysis of data already collected is
continuing but no new kaon experiments focussed on LFV are being
planned.  Interest in probing LFV has largely migrated to the muon
sector.

\section{One loop decays}

\begin{figure}[h]
 \includegraphics[angle=90, height=.3\textheight]{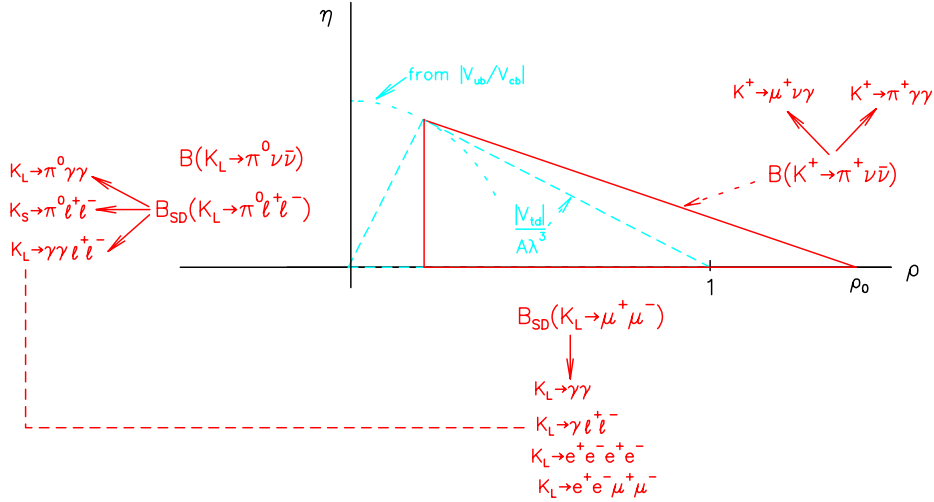}
  \caption{
$K$ decays and the unitarity plane.  The usual unitarity triangle
is dashed.  The triangle that can be constructed from rare $K$ decays
is solid.  See text for further details.
    \label{fig:triangle} }
\end{figure}

In the kaon sector the focus of experimental effort has shifted from LFV to
``one-loop'' decays.  These are GIM-suppressed decays in which loops
containing weak bosons and heavy quarks dominate or at least
contribute measurably to the rate.  These processes include \kpnn0,
\kpnnp, \kmm, $K_L \to \pi^0 e^+ e^-$ and $K_L \to \pi^0\mu^+\mu^-$.
In some cases the one-loop contributions violate CP. In one, \kpnn0,
this contribution completely dominates the decay\cite{Littenberg:1989ix}.  
Since the GIM-mechanism enhances the contribution of heavy quarks 
in the loops, in the SM these decays are sensitive to the product of couplings
$V_{ts}^* V_{td}$, often abbreviated as $\lambda_t$. Although one can 
write the branching ratio for these decays in terms of the real and 
imaginary parts of $\lambda_t$\cite{Littenberg:2000jn}, for comparison with 
other results such as those from the $B$ system,
it is convenient to express them in terms of the Wolfenstein parameters,
$A$, $\rho$, and $\eta$.  Fig.~\ref{fig:triangle} relates 
rare kaon decays to the unitarity triangle.  The dashed triangle 
is the usual one derived from $V^*_{ub} V_{ud} + V^*_{cb} V_{cd} +
V^*_{tb} V_{td}  = 0$, whereas the solid triangle illustrates the information
available from rare kaon decays.   Note that the 
apex, $(\rho, \eta)$, can be determined from either triangle, and
disagreement between the $K$ and $B$ determinations implies physics beyond 
the SM. In Fig.~\ref{fig:triangle} 
the branching ratio closest to each side of the solid triangle can be used to
determine the length of that side.  The arrows leading outward from those
branching ratios point to processes that need to be studied either because
they potentially constitute backgrounds, or because knowledge of them is
required to relate the innermost branching ratios to the lengths of
the triangle sides.
$K_L \to \mu^+ \mu^-$, which can determine the bottom of the triangle ($\rho$),
is the process for which the experimental data is the best, but for which
the theory is most problematical.  \kpnn0, which determines
the height of the triangle is theoretically the cleanest, but for
this mode experiment is many orders of magnitude short of the SM-predicted 
level.  \kpnnp, which determines the hypotenuse, is nearly as clean as \kpnn0~
and has been observed. 
Prospects for \kpnnp~are probably the best of the three since it
is clean and it is already clear it can be exploited.

\subsection{$K_L \to \mu^+ \mu^-$}

\begin{figure}[h]
 \includegraphics[angle=90, height=.075\textheight]{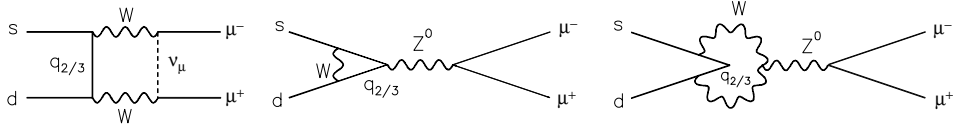}
  \caption{
Short distance contribution to $K_L \to \mu^+ \mu^-$ in the SM.
    \label{fig:mmloops} }
\end{figure}

\begin{figure}[h]
 \includegraphics[angle=0, height=.10\textheight]{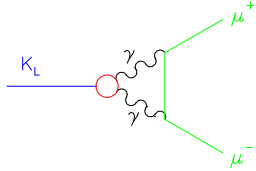}
  \caption{
Long distance contribution to  $K_L \to \mu^+ \mu^-$.
    \label{fig:mulong} }
\end{figure}

The short distance component of this decay, which arises out of the
diagrams shown in Fig.~\ref{fig:mmloops}, can be  quite reliably
calculated in the SM\cite{Buchalla:1994wq}.  The most recent
measurement of its branching ratio\cite{Ambrose:2000gj} based on $\sim 6200$
events gave $B(K_L \to \mu^+ \mu^-) = (7.18 \pm 0.17)\times 10^{-9}$.
However \kmm~ is dominated by long distance effects, the
largest of which, the absorptive contribution mediated by $K_L \to
\gamma\gamma$ shown in Fig.~\ref{fig:mulong}, accounts for 
$(7.07 \pm 0.18)\times 10^{-9}$.

Subtracting the two, yields a 90\% CL upper limit on the total
dispersive part of \bkmm~ of $0.37 \times 10^{-9}$.  One can do
somewhat better than this in the following way.  The actual quantity
measured in Ref~\cite{Ambrose:2000gj} was $\bkmmr /\bkppr =
(3.48 \pm 0.05)\times 10^{-6}$.  It is necessary to subtract from this
measured quantity the ratio $B^{abs}_{\gamma\gamma}(K_L \to
\mu^+\mu^-) \bkppr$.  Eqn~\ref{foo} shows the components of
this latter ratio, obtained from Ref.~\cite{Hagiwara:2002pw}
augmented by a new
measurement $\Gamma(K_S \to \pi^+ \pi^-)/\Gamma(K_S \to \pi^0 \pi^0)= 2.236 \pm 0.003
\pm 0.015$~\cite{Aloisio:2002bs} , whose product is $(3.344 \pm 0.053)
\times 10^{-6}$.

\bea
{\rm recently~measured}~~~~~~~~~~~~~~~~~~~~ \cr
{\rm by~KLOE}~~~~~~~~~~~~~~~~~~~~~~~~~ \cr
{\rm last~measured~by}~~~~~~~~~~~|~~~~~~~~~~~~~~~~~~~~~~~~~~~~~~~~\cr
{\rm calculated}~~~~~~~~{\rm NA31~in~1987}~~~~~~~~~~~~|~\frac{B(K_L \to \pi^0 \pi^0)}{B(K_L \to \pi^+ \pi^-)}~~~~~ \cr
|~~~~~~~~~~~~~~~~~~~~~~|~~~~~~~~~~~~~~~~~~~~~~~~|~~~~~~~~~|~~~~~~~~~~~~~~~~~~~~~~ \cr
\downarrow\,~~~~~~~~~~~~~~~~~~~\downarrow\,~~~~~~~~~~~~~~~~~~~~~\downarrow\,~~~~~~~|~~~~~~~~~~~~~~~~~~~~~~ \cr
\frac{B^{obs}_{\gamma\gamma}(K_L \to \mu\mu)}{B(K_L \to \pi^+ \pi^-)} =
\frac{B^{obs}_{\gamma\gamma}(K_L \to \mu\mu)}{B(K_L \to \gamma\gamma)}
\frac{B(K_L \to \gamma\gamma)}{B(K_L \to \pi^0 \pi^0)} 
\overbrace{\frac{B(K_S \to \pi^0 \pi^0 )}{B(K_S \to \pi^+ \pi^-)}
(1 - 6 Re \frac{\epsilon'}{\epsilon})} \cr
|~~~~~~~~~~~~~~~~~~~~~~~|~~~~~~~~~~~~~~~~~~~~~~~~|~~~~~~~~~~~~~~~~~~~~~|~~~~~~~~~~ \cr
1.195 \cdot 10^{-5}~~~~~~~~~~~~~~|~~~~~~~~~~~~~~~~~~~~~~~~|~~~~~~~~~~~~~~~~~~~~~|~~~~~~~~~~ \cr
0.632 \pm 0.009~~~~~~~~~~~~~~|~~~~~~~~~~~~~~~~~~~~~|~~~~~~~~~~ \cr
(2.236 \pm 0.015)^{-1}~~~~~~~~|~~~~~~~~~~ \cr
1-6(16.6 \pm 1.6)\cdot 10^{-4} \cr
\label{foo}
\eea

The subtraction yields $\frac{B^{disp}(K_L \to \mu^+ \mu^-)}{\bkppr} = 
(0.136 \pm 0.073) \times 10^{-6}$ (where $B^{disp}$ refers to the 
dispersive part of \bkmm).  $\frac{B^{disp}(K_L \to \mu^+ \mu^-)}{\bkppr}$
can then be multiplied by $\bkppr = (2.084 \pm 0.032) \times 10^{-3}$
~\cite{Hagiwara:2002pw} to obtain
$B^{disp}(K_L \to \mu^+\mu^-) = (0.283\pm 0.151) \times 10^{-9}$, or
$B^{disp}(K_L \to \mu^+\mu^-) < 0.47 \times 10^{-9}$ at 90\% CL.  Note
that some of the components represent 15 year-old measurements.  Since
\bkmm~and $B^{abs}_{\gamma\gamma}(K_L \to \mu^+ \mu^-)$ are so close,
small shifts in the component values could have relatively large consequences
for  $B^{disp}(K_L \to \mu^+\mu^-)$.  Several of the components could be
remeasured by experiments presently in progress.
Now if one inserts
the result of even very conservative recent CKM fits into the formula for 
the short distance part of \bkmm, one gets poor agreement with
the limit of $B^{disp}(K_L \to \mu^+\mu^-)$ derived above.  For example the
95\% CL fit of Hocker et al.\cite{Hocker:2001xe,Hocker:2001jb}, $\bar\rho = 0.07 - 0.37$,
gives $B^{SD}(K_L \to \mu^+\mu^-) = (0.4-1.3)\times 10^{-9}$.  So why 
haven't we been hearing about this apparent violation of the SM?  There
are certainly viable candidates for BSM contributions to this 
decay~\cite{Isidori:2002qe,D'Ambrosio:2002fa}.

\begin{figure}[t]
 \includegraphics[angle=0, height=.5\textheight]{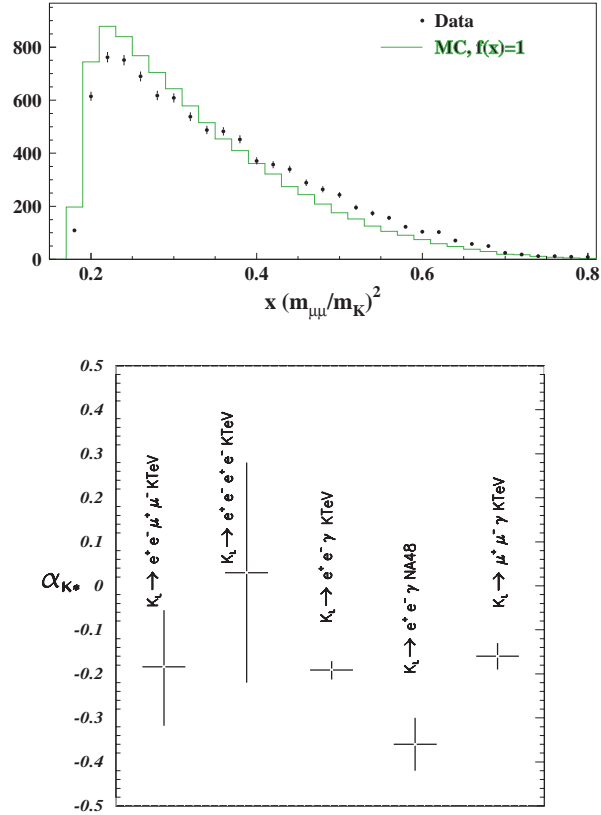}
  \caption{{\bf Top:} spectrum of $x = (m_{\mu\mu}/m_K)^2$ in 
$K_L \to \mu^+ \mu^- \gamma$ from Ref.~\protect\cite{Alavi-Harati:2001wd}.  
{\bf Bottom:} determinations of the BMS parameter
$\alpha_{K^*}$ from four $K_L$ decays involving virtual photons.
    \label{fig:mmg} }
\end{figure}

The answer is that unfortunately $K_L \to \gamma^* \gamma^* $ also
gives rise to a dispersive contribution, that is much less tractable
than the absorptive part, and which can interfere with the
short-distance weak contribution that one is trying to extract.  The
problem in calculating this contribution is the necessity of including
intermediate states with virtual photons of all effective masses.
Thus such calculations can only be partially validated by studies of
kaon decays containing virtual photons in the final state.  The
degree to which this validation is possible is controversial with
both optimistic~\cite{D'Ambrosio:1998jp,GomezDumm:1998gw}
and pessimistic~\cite{Valencia:1998xe,Knecht:1999gb} conclusions
available.  Recently
there have been talks and publications on $K_L \to \gamma e^+ e^-$
\cite{LaDue:2002} (93,400 events), $K_L \to \gamma \mu^+
\mu^-$\cite{Alavi-Harati:2001wd} (9327 events), $K_L \to e^+ e^- e^+
e^-$\cite{Alavi-Harati:2001ab} (441 events), and $K_L \to \mu^+ \mu^-
e^+ e^-$\cite{Hamm:2002vy} (133 events) and there exist
slightly older high statistics data on $K_L \to \gamma e^+
e^-$\cite{Fanti:1999rz} (6864 events).  Figure~\ref{fig:mmg}-top
shows the spectrum of $x = (m_{\mu\mu}/m_K)^2$ from
Ref.\cite{Alavi-Harati:2001wd}.  The disagreement between the data
(filled circles with error bars) and the prediction of pointlike
behavior (histogram) clearly indicates the presence of a form factor.  A
long-standing candidate for this is provided by the BMS
model\cite{Bergstrom:1983rj} which depends on a single parameter,
$\alpha_{K^*}$.

Fig.~\ref{fig:mmg}-bottom shows five determinations of this parameter.
The KTeV data are internally consistent, but there's a disagreement
with the NA48 result for $K_L \to e^+ e^- \gamma$~\cite{Fanti:1999rz}
Fitting to the more recent DIP parameterization of these
decays~\cite{D'Ambrosio:1998jp} gives a similar level of agreement.
Both these parameterizations predict a connection between the shape
of the $\ell^+ \ell^-$ spectra and the branching ratios which can
be exploited in the cases involving muon pairs.  In those cases 
the parameters have been determined from both the spectra and the
branching ratios.  These determinations agree at the 1-2$\,\sigma$ level.
However the available data is not yet sufficient to clearly favor either
parameterization.  In addition, it seems clear that a very large increase
in the data of processes where both photons are virtual, such as 
$K_L \to e^+ e^- \mu^+ \mu^-$ would be needed to allow a real test of the 
DIP parameterization~\cite{Hamm:2002vy,Halkiadakis:2001fb}.
Additional effort, both experimental and theoretical, is required
before the quite precise data on \bkmm~ can be fully exploited.

\begin{figure}[t]
 \includegraphics[angle=0, height=.27\textheight]{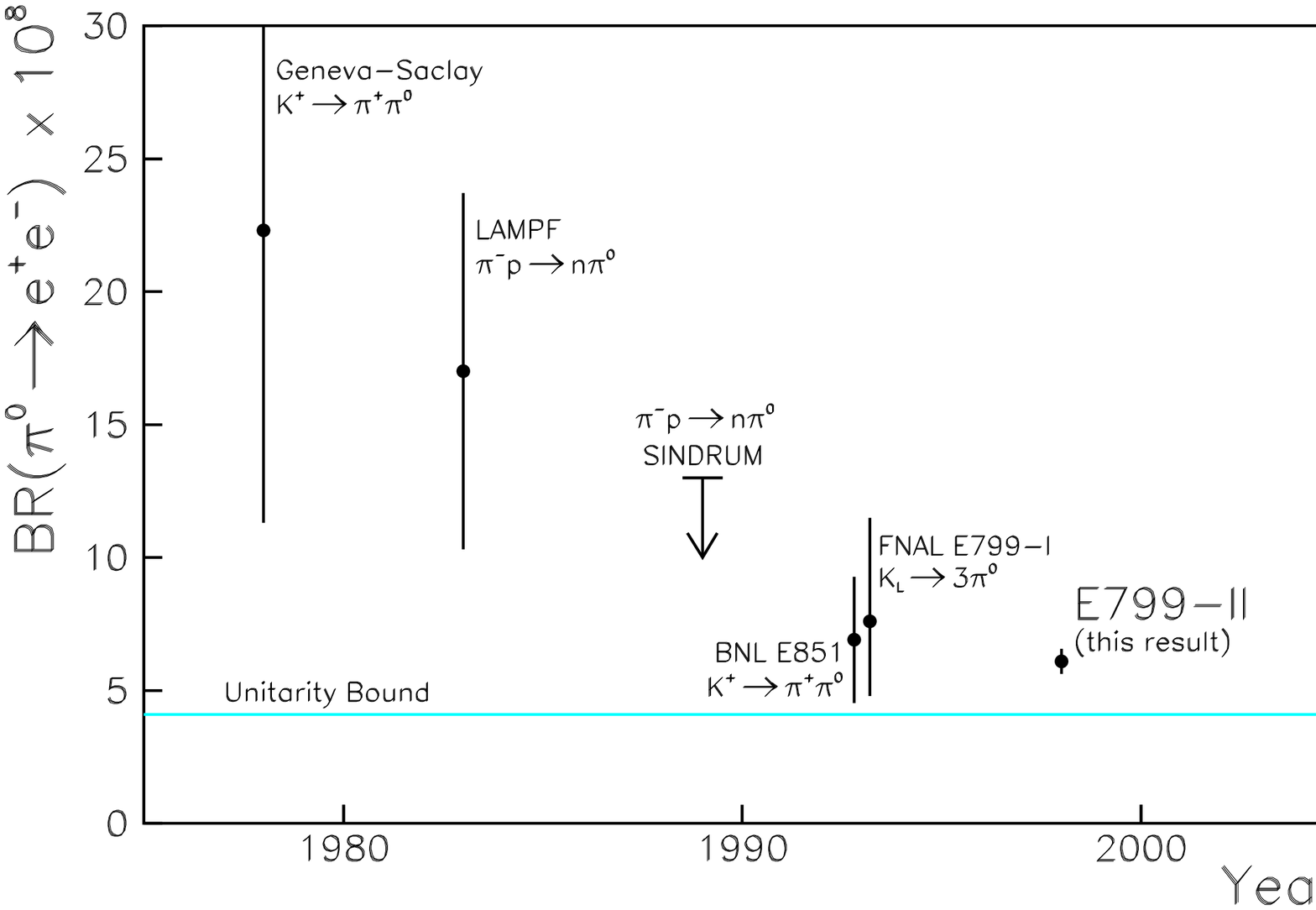}
 \includegraphics[angle=0, height=.27\textheight]{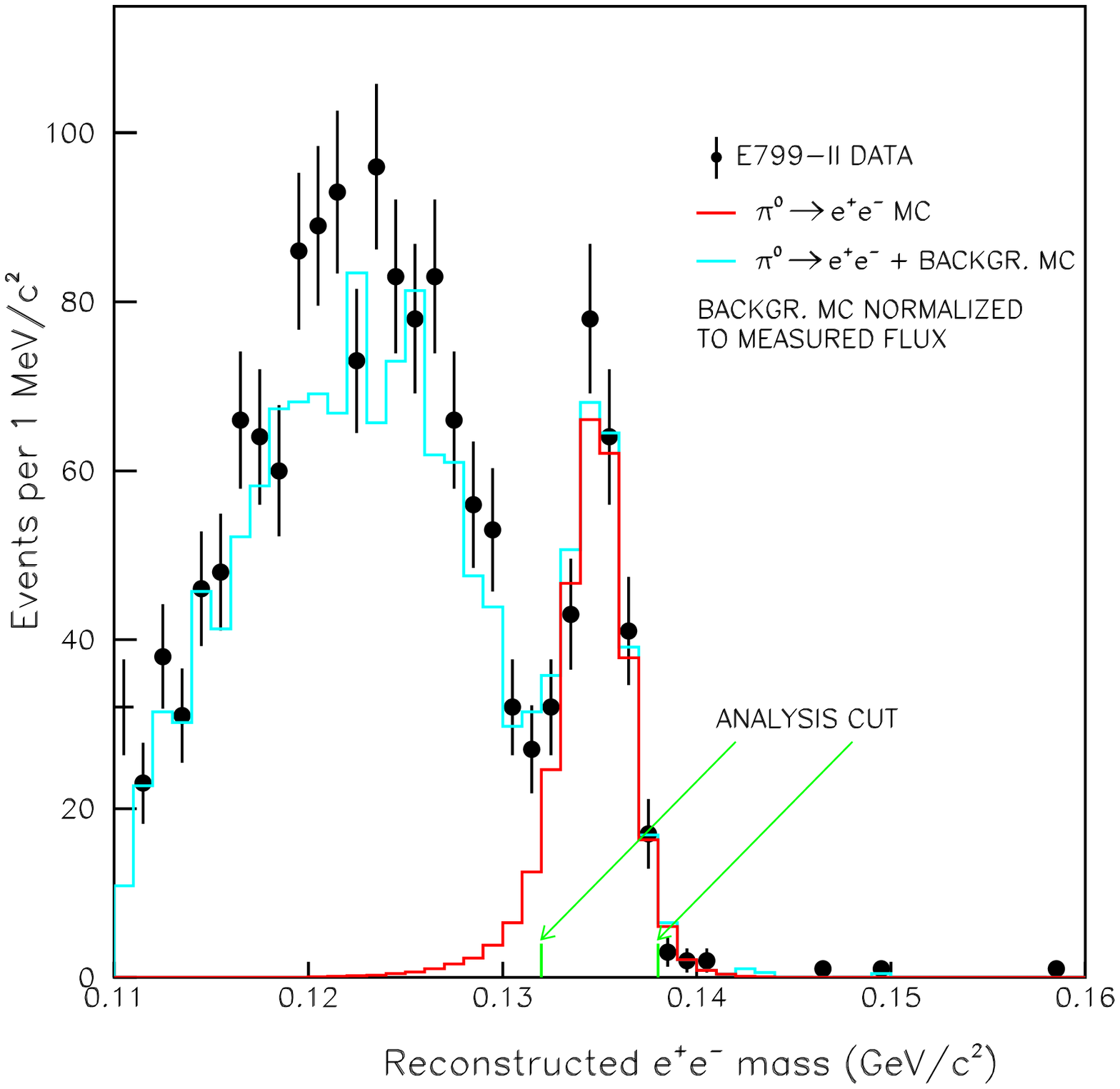}
  \caption{{\bf Left:} history of $\pi^0 \to e^+ e^-$ measurements
from Ref.~\protect\cite{Zimmerman:1999}.  Horizontal line is the absorptive
contribution from $\pi^0 \to \gamma \gamma$.
{\bf Right:} Reconstructed $m_{ee}$ distribution for $\pi^0 \to e^+ e^-$ candidates
from Ref.~\protect\cite{Zimmerman:1999}.  
    \label{fig:pi0ee} }
\end{figure}

Another possible avenue to understanding long-distance dispersive
effects in $K_L \to \mu^+ \mu^-$ may be afforded by the study of
$\pi^0 \to e^+ e^-$ where similar effects can come into 
play\cite{Knecht:1999gb}.  This
subject has had somewhat of a checkered history, as indicated in
Fig.~\ref{fig:pi0ee} left, but a 1999 KTeV result in which the $\pi^0$'s
were tagged in $K_L \to 3 \pi^0$ decays has definitively established
the presence of a dispersive term~\cite{Alavi-Harati:1999zr}.  The clear signal
from this experiment is shown in Fig.~\ref{fig:pi0ee} right.  Hopefully
theorists will now find this result a useful test-bed.

	Finally, one might ask if it is possible to extract short distance
information from the decay $K_L \to e^+ e^-$.  AGS E871 has seen four events
of this mode, establishing a branching ratio of $(8.7 {+5.7 \atop{-4.1}})
\times 10^{-12}$~\cite{Ambrose:1998cc}.  This is the smallest elementary 
particle branching ratio yet measured.  Unfortunately the SM short
distance contribution is helicity suppressed with respect to
$K_L \to \mu^+ \mu^-$ by the ratio
$m_e^2/m_{\mu}^2$ while the suppression of long distance contributions
is mitigated by logarithmic enhancements, making the extraction of SM
short distance information almost impossible.  Ironically, the dispersive
long-distance contribution can be reliably calculated in this
case~\cite{Valencia:1998xe}.  The theoretical prediction of the
branching ratio agrees well with what is observed, which limits the
presence of BSM pseudoscalar couplings in this decay.

\subsection{\kpnnp}

From the point of view of theory \kpnnp~ is remarkably clean, suffering 
none of the 
long distance complications to which $K_L \to \mu^+ \mu^-$ is subject.
The often problematical hadronic matrix element
can be calculated to $\sim 2\%$ via an isospin
transformation from that of $K_{e3}$\cite{Marciano:1996wy}.
\kpnnp~ is very 
sensitive to $V_{td}$ (it is actually directly sensitive to the quantity
$V_{ts}^* V_{td}$).  Its amplitude is proportional to
the hypotenuse of the solid triangle in Fig.~\ref{fig:triangle}.  This is
equal to the vector sum of the line proportional to $V_{td}/A \lambda^3$
(where $\lambda \equiv sin \theta_{Cabibbo}$)
and that from $(1,0)$ to the point marked $\rho_0$. The length $\rho_0 -1$ 
along the real
axis is proportional to the amplitude for the charm contribution to
\kpnnp.  The QCD corrections to this amplitude, which
are responsible for the largest uncertainty in 
\bkpnnp, have been calculated to NLLA\cite{Buchalla:1994wq}. The residual
uncertainty in the charm amplitude is estimated to be $\sim 15\%$ which
leads to only  a $\sim 6\%$ uncertainty\cite{Buchalla:1998ba} in
extracting $|V_{td}|$ from \bkpnnp.  

\begin{figure}[h]
\includegraphics[angle=0, height=.45\textheight]{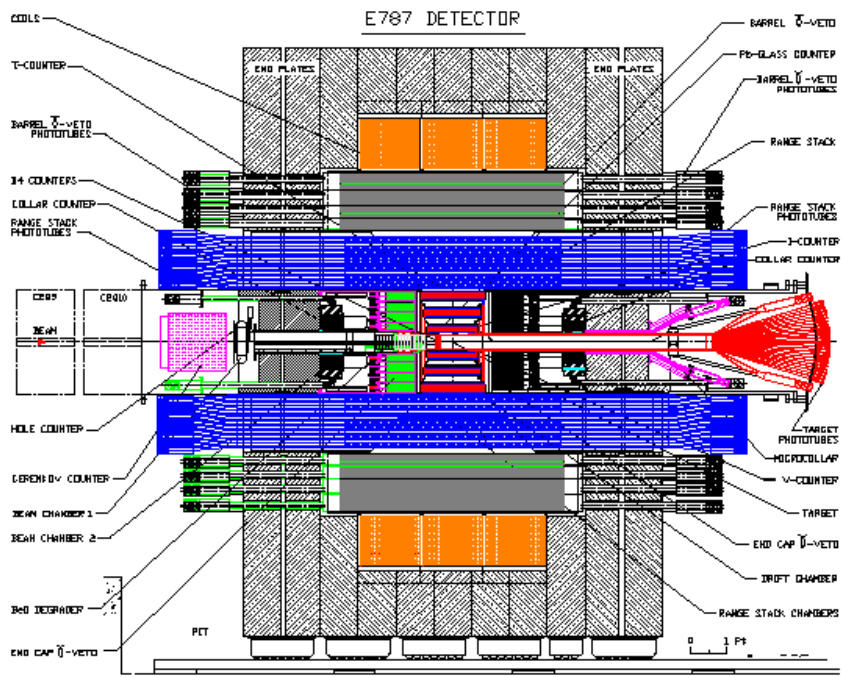}
  \caption{
E787 detector.
    \label{fig:det787} }
\end{figure}

	Like all previous experiments AGS E787 worked with stopped
$K^+$.  This gives direct access to the $K^+$ center of mass, and
facilitates hermetic vetoing.  The detector, shown in
Fig.~\ref{fig:det787} was a cylindrically symmetric solenoid inside a
1 Tesla solenoid~\cite{Atiya:1992vh}.  It was situated at the end of
a $\sim 700$ MeV/c electrostatically separated beam that provided an
80\% pure supply of $> 10^7~K^+$ per AGS cycle\cite{Doornbos:2000hb}.
Beam particles traversed a Cerenkov counter that identified $K^+$ and
$\pi^+$ and were tracked through two stations of MWPC's. They were
then slowed in a BeO degrader followed by a lead glass photon veto.
Approximately one quarter of them survived to exit the lead glass and 
traverse a hodoscope before entering a scintillating fiber stopping
target.   A hodoscope surrounding the stopping target was used
to trigger on a single charged particle leaving the target after a delay
of $\sim 0.12\, \tau_K$.  The particle was subsequently tracked in a
low-mass
cylindrical drift chamber allowing its momentum to be precisely
determined.  Additional trigger counters required the particle to exit the
chamber radially outward and enter a cylindrical array of
scintillators and straw chambers (the Range Stack) in which it was required
to stop in order to be considered a $K^+ \to \pi^+ \nu\bar\nu$ candidate.
The Range Stack scintillation counters were read out by
phototubes on both ends which allowed a determination of the position
of tracks in the beam direction via differential timing and pulse
height.  This facility, along with the pattern of pulse heights excited
in the  counters and the coordinates measured in two layers of straw chambers,
determined the range of the stopping particles.  The
detector design minimized ``dead'' material so that the
kinetic energy could also be well measured.  Range/energy/momentum
comparison is a powerful discriminator of low energy particle identity.
In addition, transient digitizer readout of the Range Stack
photomultipliers allowed the $\pi^+ \to \mu^+ \to e^+$ decay chain to
be used to identify $\pi^+$'s.  The combination of kinematic and
life-cycle techniques can distinguish pions from muons with a
misidentification rate of $\cal{O}$$(10^{-8})$.  Surrounding the Range
Stack was a cylindrical lead-scintillator veto counter array and
adjacent to the ends of the drift chamber were endcap photon veto arrays
of CsI-pure modules\cite{Chiang:1995ar}.  There were also a number of
auxillary veto counters near the beamline.  

        Monte Carlo estimation of backgrounds was in general not
reliable since it was necessary to estimate rejection factors as high
as $10^{11}$ for decays occurring in the stopping
target.  Instead, methods were developed to measure the background from
the data itself, using both the primary data stream and data from
special triggers taken simultaneously.  The principles developed
included:

\begin{itemize}
\item To eliminate bias $-$ the signal acceptance region is kept hidden while
cuts are developed.

\item Cuts are developed on 1/3 of the data but residual
background levels are measured on the remaining 2/3.

\item Bifurcated background calculation.  Background sources are identified
{\it a priori}.  Two independent cuts with high individual rejection are
developed for each background.  Each cut is reversed in turn as the other
is studied.  After optimization, the combined effect of the cuts can then be
calculated as a product.

\item Cuts are loosened to uncover correlations.  If any are found,
they are applied before the bifurcation instead of after it, and the
background determination process repeated.

\item Background calculations are checked by comparison with
data near the signal region.

\end{itemize}

In this way backgrounds can be reliably calculated at the $10^{-3}$
to $10^{-2}$ event level.

        All factors in the acceptance besides those of solid angle,
trigger and momentum interval were determined from data.

\begin{figure}[t]
\includegraphics[angle=0, height=.275\textheight]{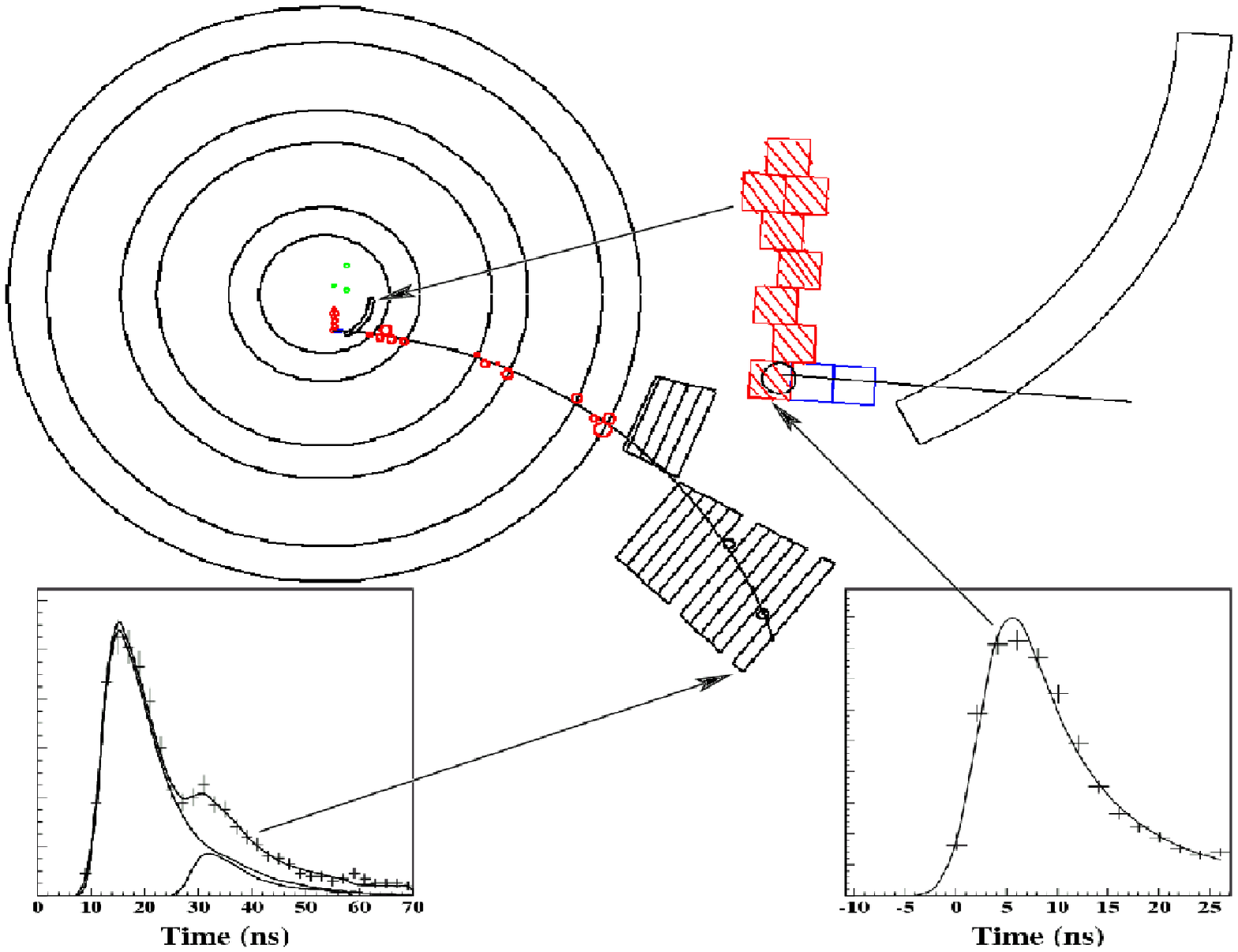}
\includegraphics[angle=0, height=.275\textheight]{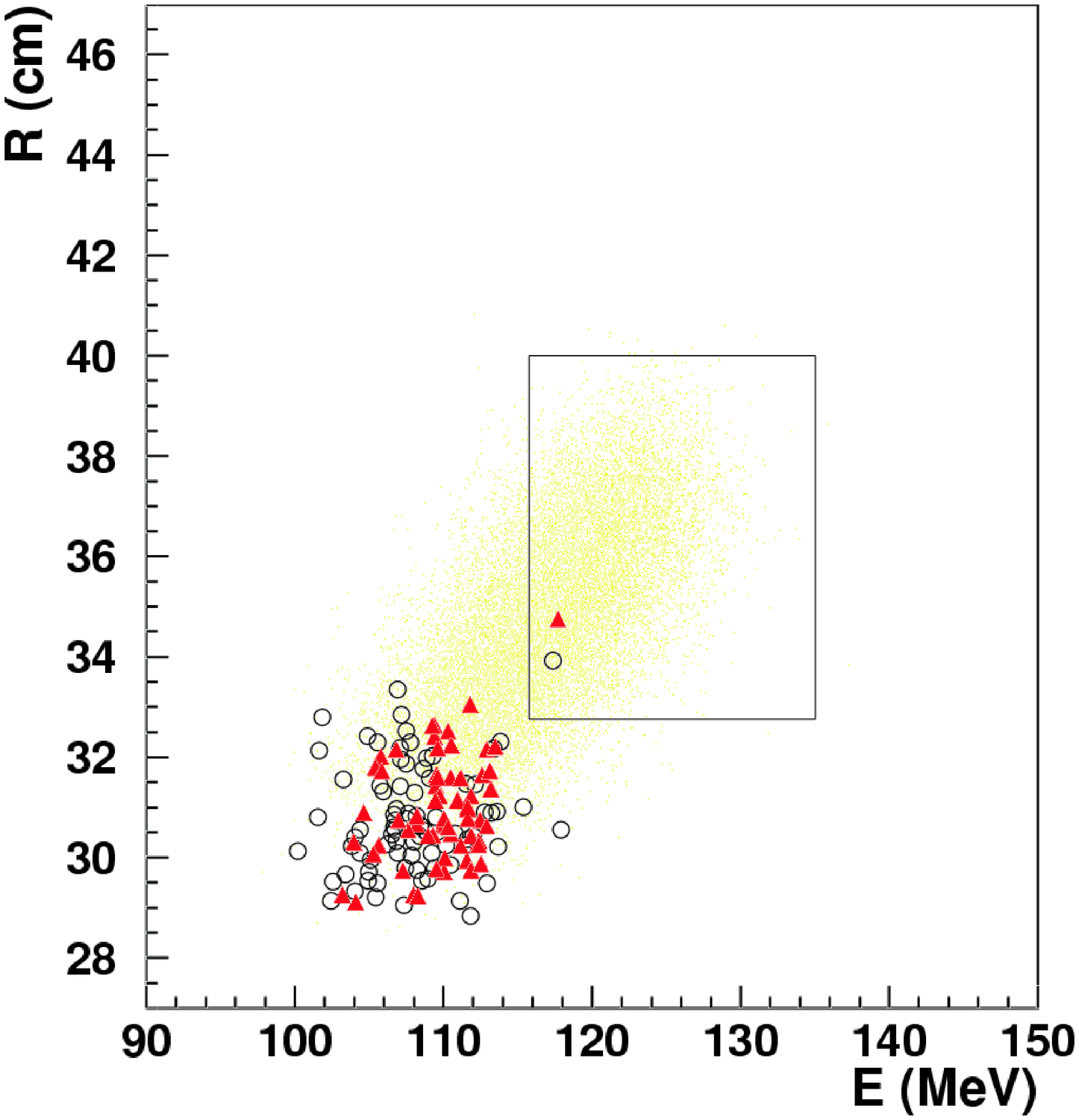}
  \caption{
{\bf Left:} new \kpnnp~event. {\bf Right:} Range vs energy of $\pi^+$ in the final
sample.  The circles are 1998 data and the triangles 1995-7 data.  The events 
around $E=108$ MeV are $K^+ \to \pi^+ \pi^0$ background.  The simulated
distribution of expected signal events is indicated by dots.
    \label{fig:e787} }
\end{figure}

	Evidence for $K^+ \to \pi^+ \nu\bar\nu$ in the form of a very
clean single event candidate was found in data taken in
1995\cite{Adler:1997am}.  Subsequent runs in 1996 and 1997 did not
produce additional events but in data collected in 1998, a second
event was found\cite{Adler:2001xv} (see Fig.~\ref{fig:e787}).
Combined with previous data \cite{Adler:2000by}, this yields a
branching ratio \bkpnnp $= (1.57 {+1.75 \atop -0.82}) \times
10^{-10}$. By comparison, a fit to the CKM phenomenology yields the
expectation $(0.72 \pm 0.21) \times 10^{-10}$\cite{D'Ambrosio:2001zh}.
The total background to the two events was measured to be 0.15 of an
event {\it i.e.} $\sim 20\%$ of the signal branching ratio predicted
by the SM.  Thus E787 has developed methods to reduce the backgrounds
to a level sufficient to make a precise measurement of \kpnnp, a 
fact that helped inspire the successor experiment described below.

	It is possible to use the E787 result to extract CKM information
under various assumptions.  One can obtain 
\begin{eqnarray}
0.007 < & |V_{td}| &< 0.030~~~~ (68\% ~ CL) \\
0.005< & |V_{td}|& ~~~~~~~~~~~~~~~~~(90\% ~ CL) \\
&|V_{td}|& < 0.033~~~~ (90\% ~ CL) \\
-0.022 < & Re V_{td}& < 0.030~~~~ (68\% ~ CL) \\
&|Im V_{td}|& < 0.028 ~~~~ (90\% ~ CL)
\label{vtd787}
\end{eqnarray}
assuming that $\overline m_t(m_t) = 166 \pm 5\,$GeV/c and 
$V_{cb} = 0.041 \pm 0.002$.  Alternatively one can extract limits on 
$\lambda_t$ that don't depend on $V_{cb}$: 
\begin{eqnarray}
0.29 < &|\lambda_t|/10^{-3} &   < 1.2 ~~~~ (68\% ~ CL) \\
-0.88< & Re(\lambda_t)/10^{-3}& < 1.2  ~~~~ (68\% ~ CL) \\
&Im(\lambda_t)/10^{-3}& < 1.1 ~~~~ (90\% ~ CL) 
\label{lt787}
\end{eqnarray}
These limits are not
competitive with what can be obtained using the full array of available 
phenomenological information, but they depend on far fewer assumptions.
It will be very  interesting to compare the large value for $|V_{td}|$
suggested by the E787 result with the value that is eventually extracted
from $\bar B_s - B_s$ mixing when it is finally observed.  

\begin{figure}[h]
\includegraphics[angle=90., width=.35\textheight]{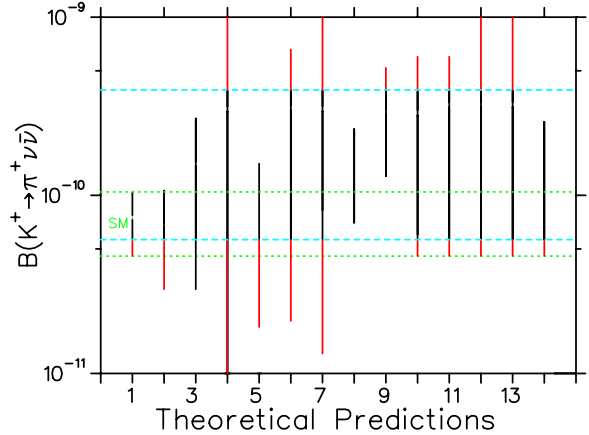}
  \caption{Predictions for $B(K^+ \to \pi^+ \nu\bar\nu)$.  The dashed 
horizontal lines indicate E787's $\pm 1 \sigma$ limits.  The dotted
horizontal lines indicate the SM range.  References are indicated in
Table\ref{tab:bsm787}
    \label{fig:bsm787} }
\end{figure}

	From the first observation published in 1997, E787's results
for $B(K^+ \to \pi^+ \nu\bar\nu)$ have been rather high with respect
to the SM prediction.  Although there has never been a statistically 
significant disagreement with the latter, this has stimulated a
number of predictions in BSM theories.  Fig~\ref{fig:bsm787} shows a
selection of such predictions (defined in Table~\ref{tab:bsm787})
compared with the SM range.

	The E787 data also yields an upper limit on
the process $K^+ \to \pi^+ X^0$ where $X^0$ is a massless weakly interacting
particle such as a familon\cite{Wilczek:1982rv}: $B(K^+ \to \pi^+ X^0)
<5.9 \times 10^{-11}$ at 90\% CL.  The case of $M_{X^0} > 0$ is discussed
below.

\begin{table}[b]
\begin{tabular}{|c|l|c|} \hline
\# & Theory & Ref. \\
\hline
1 & Standard Model & \cite{Buras:2001pn} \\
2 & MSSM with no new sources of flavor- or CP-violation & \cite{Buras:2000dm} \\
3 & Generic SUSY with minimal particle content & \cite{Buras:1999da} \\
4 & Upper limit from $Z'$ limit given by $K$ mass difference & \cite{Long:2001bc} \\
5 & Topcolor & \cite{Buchalla:1996dp} \\
6 & Topcolor-assisted Technicolor Model & \cite{Xiao:1999pt} \\
7 & Multiscale Walking Technicolor Model & \cite{Xiao:1999ps} \\
8 & $SU(2)_L \times SU(2)_R$ Higgs & \cite{Chanowitz:1999jj} \\
9 & Four generation model & \cite{Hattori:1999ap} \\
10 & Leptoquarks & \cite{Agashe:1996qm} \\
11 & R-parity-violating SUSY & \cite{Bhattacharyya:1998be} \\
12 & Extension of SM to gauge theory of $J=0$ mesons & \cite{Machet:1999dj} \\
13 & Multi Higgs Multiplet Model & \cite{Grossman:1994jb} \\
14 & Light sgoldstinos & \cite{Gorbunov:2000cz}\\
\hline
\end{tabular}
\caption{Predictions for $B(K^+ \to \pi^+ \nu\bar\nu)$} 
\label{tab:bsm787}
\end{table}

\begin{figure}[h]
\includegraphics[angle=0, height=.275\textheight]{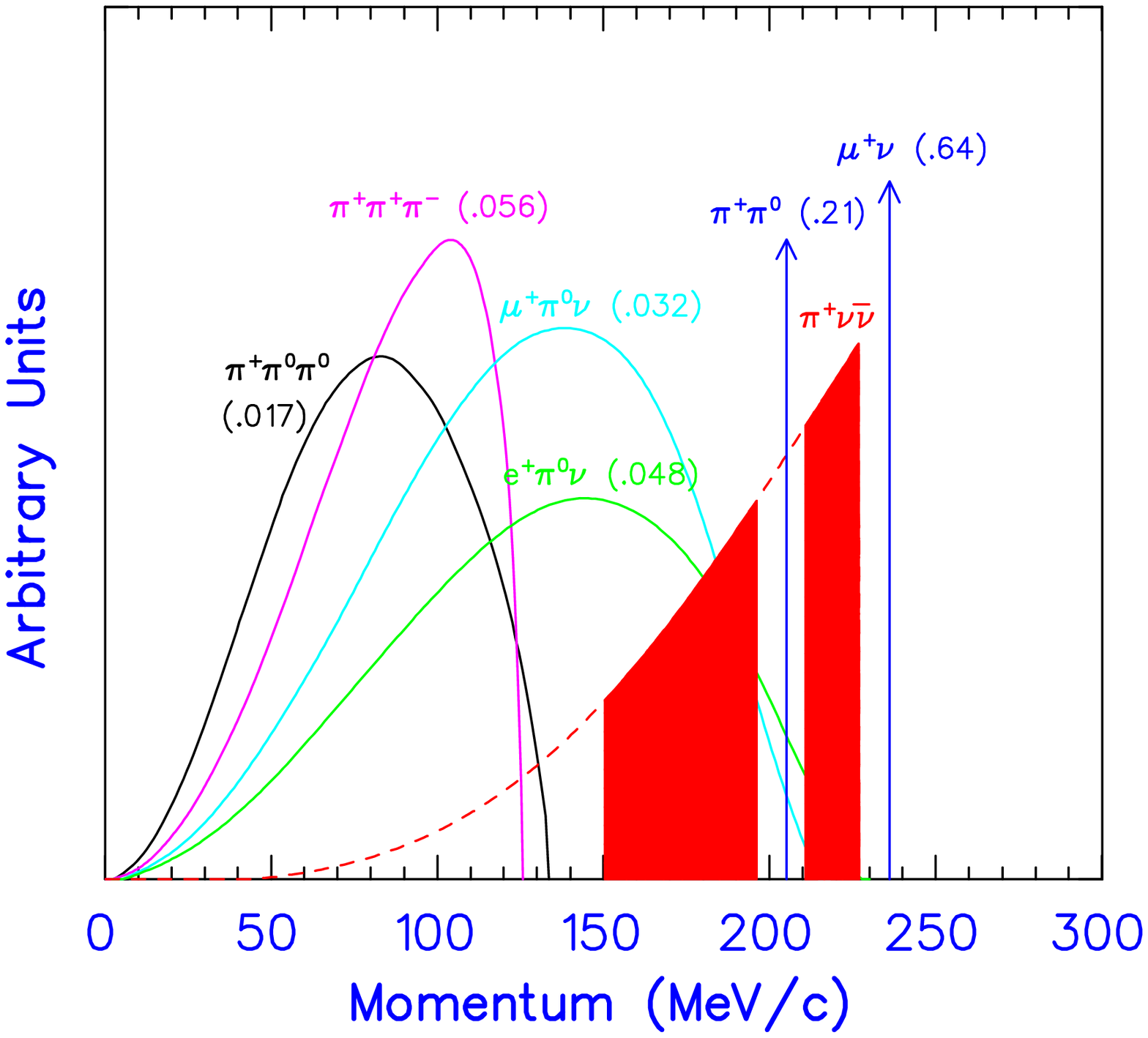}
\includegraphics[angle=0, height=.275\textheight]{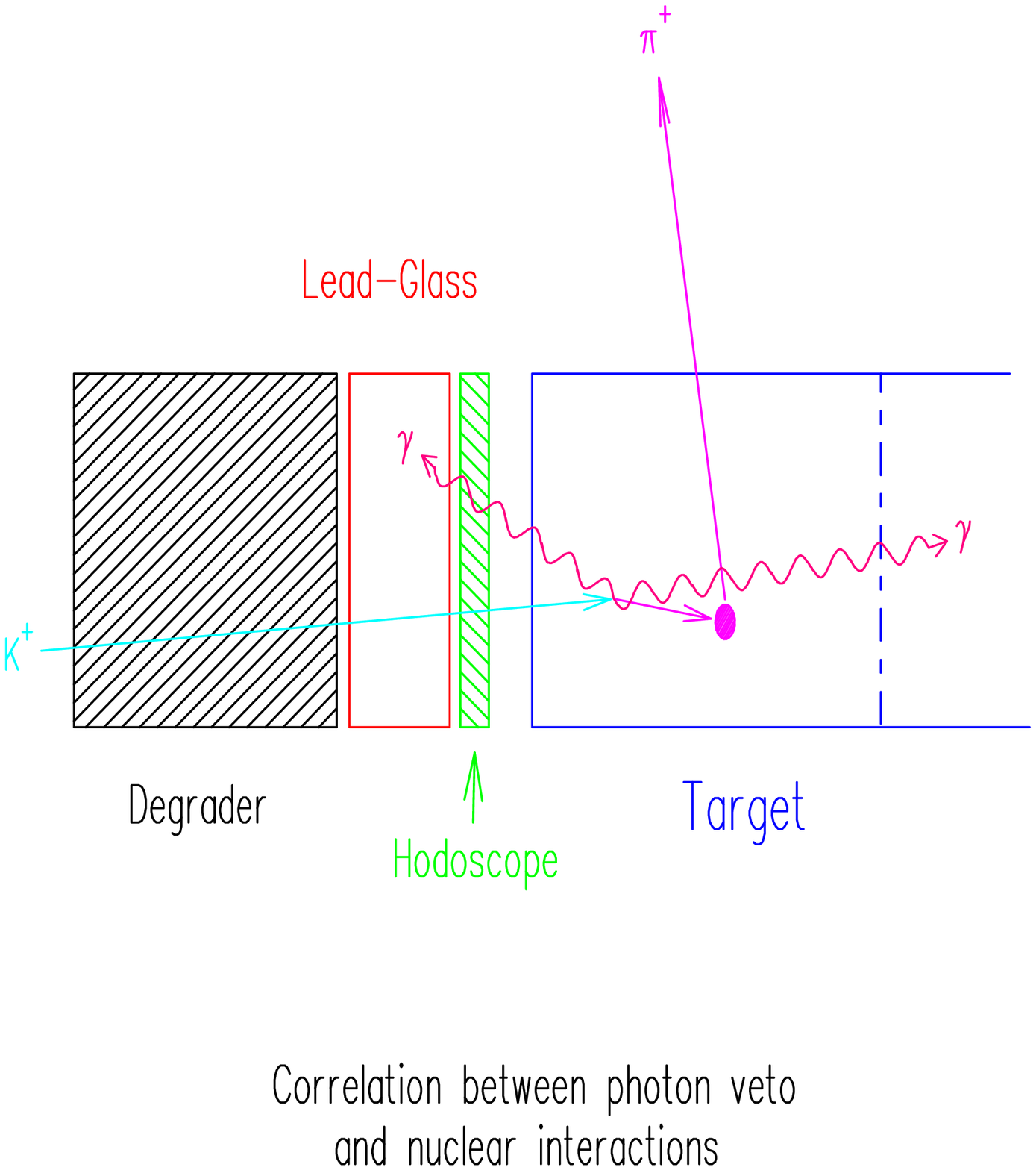}
  \caption{{\bf Left}: Center of mass momentum spectrum of $\pi^+$ from
$K^+ \to \pi^+ \nu\bar\nu$ compared with charged product spectra of the 
seven most common $K^+$ decays.  Filled areas indicate the portions
of the spectrum used in E787 analyses.  {\bf Right}: Cartoon of 
limiting background in the softer region of the $K^+ \to \pi^+ \nu\bar\nu$
spectrum.  See text for details.
    \label{fig:pnn2} }
\end{figure}

	Fig.~\ref{fig:pnn2} left shows the $\pi^+$ momentum spectrum
from $K^+ \to \pi^+ \nu\bar\nu$ in the SM, along with the charged
track spectra from other kaon decays.

\begin{figure}[h]
\includegraphics[angle=0, height=.300\textheight]{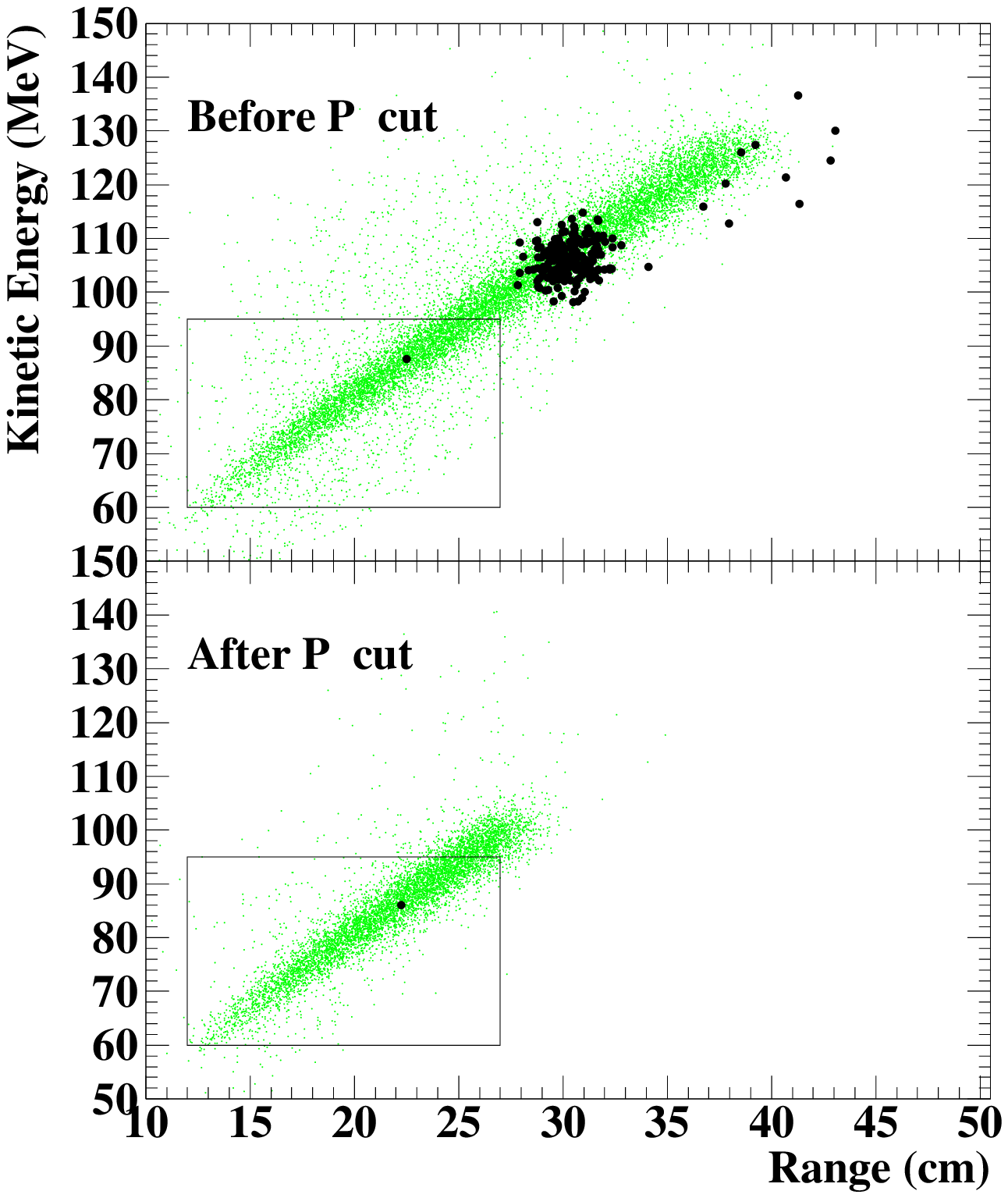}
\includegraphics[angle=0, height=.300\textheight]{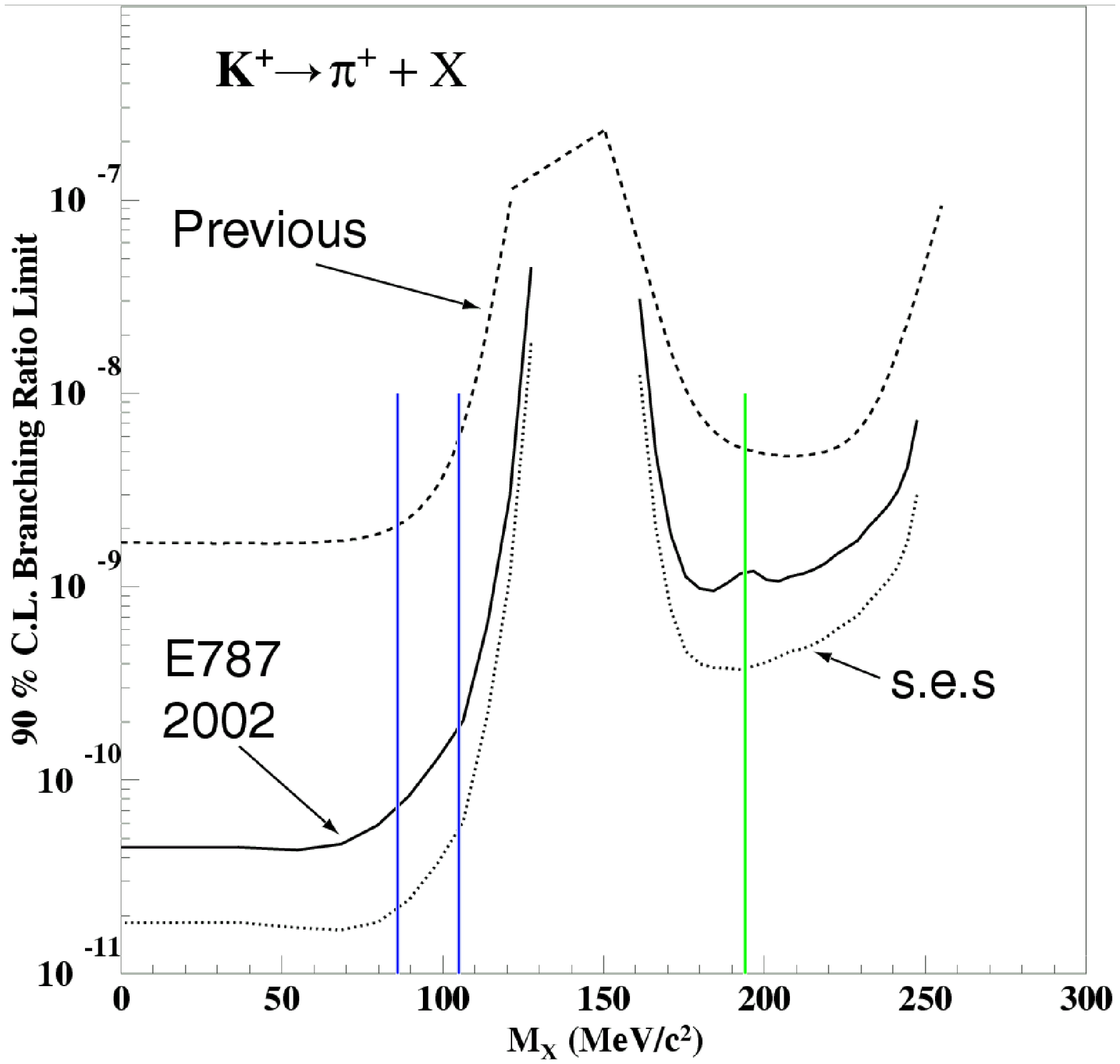}
  \caption{{\bf Left}: Signal plane for $K^+ \to \pi^+ \nu\bar\nu$
analysis of soft $\pi^+$ region. 
{\bf Right}: Limit on $K^+ \to \pi^+ X^0$ vs $m_X$.  Vertical lines
indicate location of events.  For comparison previous limits are
indicated as is the single event sensitivity limit of E787.
    \label{fig:pnn2_rslt} }
\end{figure}

	E787 is sensitive to the filled-in regions of the $K^+ \to \pi^+
\nu\bar\nu$ spectrum in Fig.~\ref{fig:pnn2}.  However all the E787 results
mentioned so far come from the region on the right, in which the
momentum of the $\pi^+$ is greater than than of the $\pi^+$ from
$K^+ \to \pi^+ \pi^0$.  The region on the left contains more of the
signal phase space, but is more subject to background from $K^+ \to \pi^+
\pi^0$.  It is relatively easy for the $\pi^+$ to lose energy through
nuclear interactions.  What is more, there is a problematical correlation
between nuclear scattering in the stopping target and the weaker
E787 photon veto in the beam region.  Fig.~\ref{fig:pnn2} right 
illustrates this problem.  A $K^+$ decays with the $\pi^+$ pointing downstream
(the $\pi^0$ must then be pointing upstream).  Normally such a decay
would not trigger the detector, but here the $\pi^+$ undergoes a 90$^\circ$
scatter, loses enough energy to get into the accepted momentum range
and heads for the drift chamber.  At the same time the $\pi^0$ decays
asymmetrically, with the high energy photon heading upstream, to where
the veto is least capable and the low energy photon heading downstream,
toward another weak veto region.  This sequence of events is unlikely,
but 20\% of $K^+$ decay to $\pi^+ \pi^0$, and one is trying to
study a process that happens one in ten billion times.  The fact that the
same scatter both down-shifts the $\pi^+$ momentum and aims the $\pi^0$
at the weak veto region confounds the usual product of rejection factors
so effective in the high momentum region.  A test analysis using 1996
data was undertaken to determine whether methods could be developed
to overcome this background.  These leaned
heavily on exploiting the transient digitized signals from the stopping
target target scintillating fibers.  At the cost of giving up some
acceptance by only using decays later than 0.5 $\tau_K$, one could
detect evidence of $\pi^+$ scattering occluded by kaon signals
in the critical target elements.  In this way a single event sensitivity
of $\sim 10^{-9}$ was achieved with a calculated background of $0.73$
events.  Fig.~\ref{fig:pnn2_rslt} left shows the resulting distribution
of $\pi^+$ kinetic energy and range for surviving candidates.
The top left shows the distribution before the final cut on $\pi^+$
momentum.  The peak at $T_{\pi} \sim 108$ MeV, $R_{\pi} \sim 30.5$ cm
is due to $K^+ \to \pi^+ \pi^0$ events.  After the final cut, one
event remains, consistent with the background estimation.  This yields
$B(K^+ \to \pi^+ \nu\bar\nu) < 4.2 \times 10^{-9}$ at 90\% CL, consistent
with other E787 results.  This kinematic region is particularly sensitive
to possible BSM effects which produce scalar or tensor pion spectra
(rather than the vector spectrum given by the SM).  One can combine
this region with the high momentum region to get 90\% CL upper limits of 
 $4.7 \times 10^{-9}$ and $2.5 \times 10^{-9}$ for scalar and tensor 
interactions,  
respectively.  These measurements are also sensitive to $K^+ \to \pi^+
X^0$ where $X^0$ is a hypothetical stable weakly interacting particle
or system of particles.  Fig.~\ref{fig:pnn2_rslt} right shows 90\% CL upper
limits on $B(K^+ \to \pi^+ X^)$ together with the previous limit from
\cite{Atiya:1993qr}.  The dotted line in the figure is the single
event sensitivity. 

A new experiment, E949\cite{e949}, based on an upgrade of the E787
detector has begun operation at the AGS.  The detector has been
improved in a number of ways with respect to E787: thicker and more
complete veto coverage, augmented beam instrumentation, higher
capacity DAQ, more efficient trigger counters, upgraded chamber
electronics, auxiliary gain monitoring systems, etc.  In addition SM
sensitivity is anticipated for the kinematic region $140 <p_{\pi^+} <
190$ MeV/c.  Several of the upgrades were aimed at exploiting this
region, and based on the test analysis discussed above, a signal/background
of 1:1 is expected for this part of the spectrum.
Using the entire flux of the AGS for 6000 hours, E949 is
designed to reach a sensitivity of $\sim 10^{-11}$/event. The
experiment made its first physics run in the Spring of 2002; the
detector operated well at fluxes nearly twice as high as those typical
of E787.

\begin{figure}[t]
 \includegraphics[angle=0, height=.375\textheight]{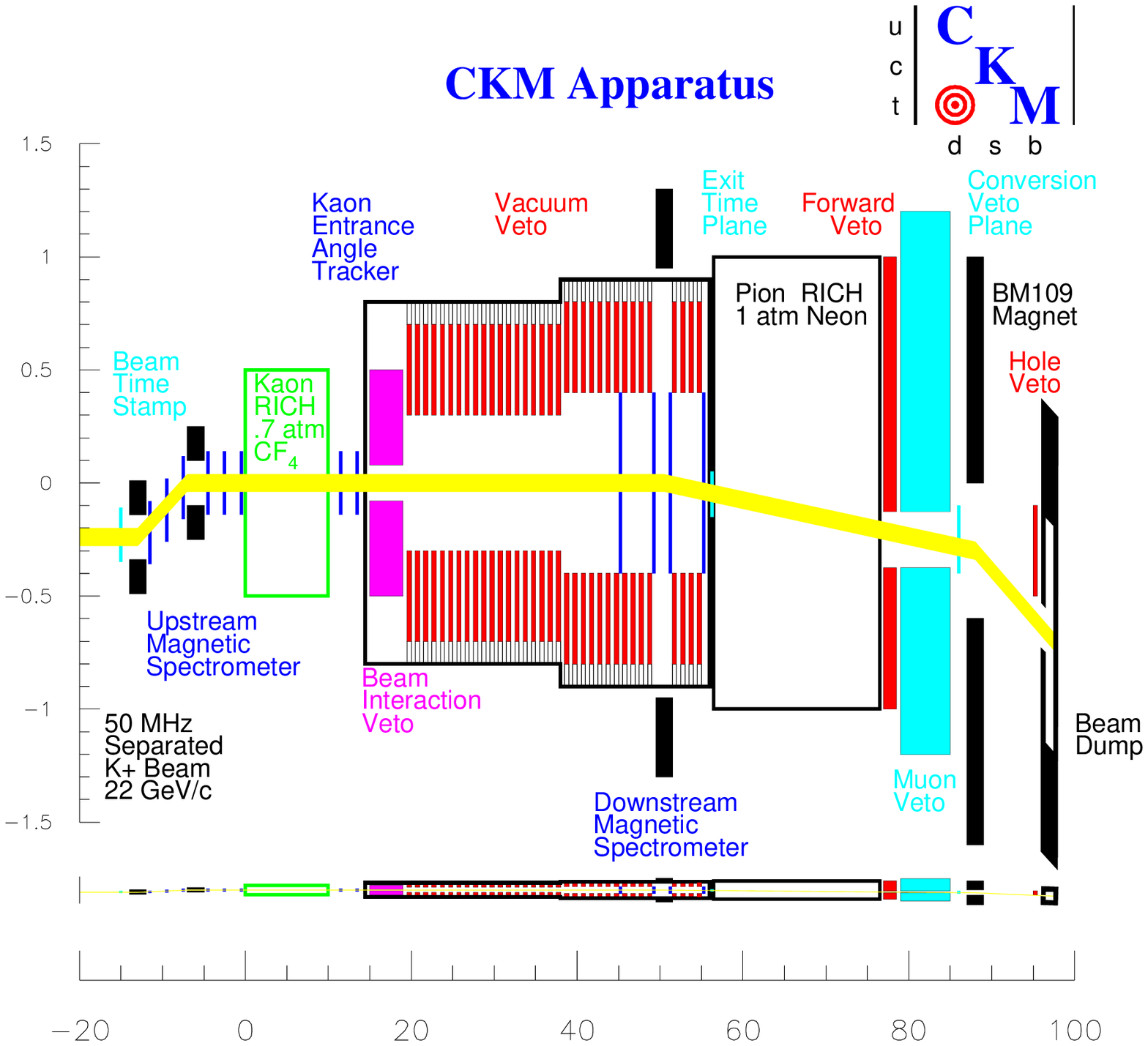}
 \includegraphics[angle=0, height=.3\textheight]{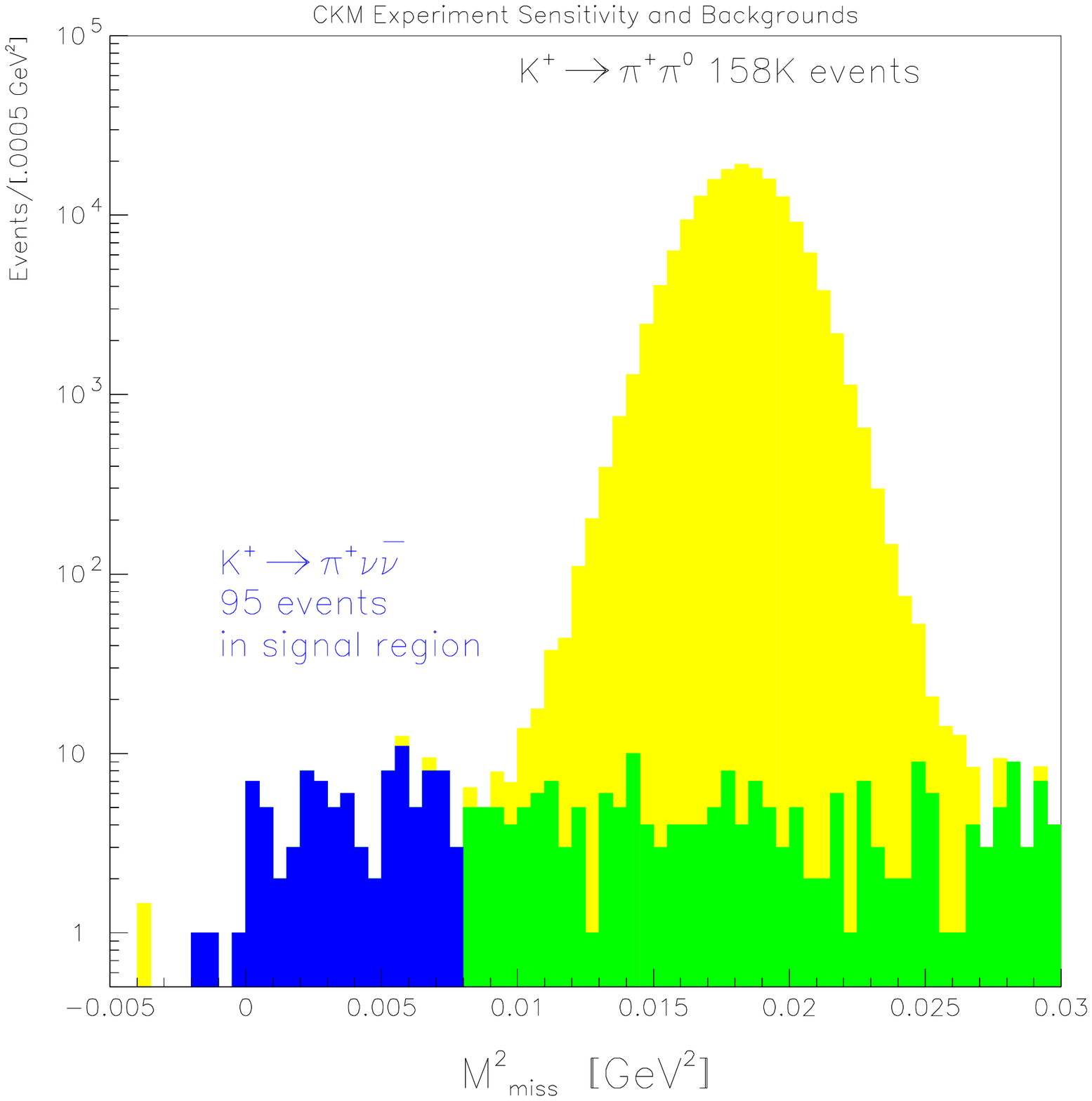}
  \caption{
{\bf Left:} apparatus of the CKM experiment at FNAL.  {\bf Right:} expected
signal and background.
    \label{fig:ckm} }
\end{figure}

In June 2001, Fermilab gave Stage 1 approval to an experiment,
CKM~\cite{Cooper:2001ba}, to extend the study of \kpnnp~by yet an
another order of magnitude in sensitivity.  This experiment, unlike
all previous ones studying this process, uses an in-flight rather than a
stopping $K^+$ technique.  Fig.~\ref{fig:ckm} shows the proposed detector.
Protons at 120 GeV/c from the Fermilab Main Injector will be used to produce
an RF-separated 22 GeV/c positive beam. The superconducting RF system
will produce a 50 MHz positive beam that is $\sim 2/3$ pure $K^+$. 
Incoming kaons 
will be momentum analyzed in a wire chamber spectrometer and
velocity analyzed by a RICH counter with phototube readout.  The RICH radiator
will be 10$\,$m of CF$_4$ at 0.7 atm.  The kaon direction will
subsequently be remeasured by a high precision tracker before entering
a vacuum decay region in which 17\% of the $K^+$ decay.
The upstream section of the decay vacuum will be surrounded by a
lead-scintillator sampling calorimeter designed to veto 
interaction products of the beam kaons in the tracking chambers.
Subsequent, wider-bore vacuum vetoes line the remainder of the decay
volume.  These will be wavelength-shifting fiber readout lead-scintillator 
sampling
calorimeters with 1mm Pb/5mm scintillator granularity and redundant
photomultiplier readout.  The system is designed to reject photons of
energy greater than 1 GeV/c with an inefficiency no greater than
$3 \times 10^{-5}$.
The decay region will be bounded by a straw-chamber-based
magnetic spectrometer, also in vacuum.   Downstream of the
spectrometer will be a second RICH counter, filled with 20m of
neon at 1 atm.  This counter is designed to tag pions and measure
their vector velocity ($\mu - \pi$ separation is 10 $\sigma$, momentum
and angular resolutions are 1\% and 200$\mu$rad respectively).
Downstream of the pion RICH will be an electromagnetic calorimeter
covering the full aperture excepting a small
hole for the beam.  This calorimeter must be highly segmented
transversely in order to allow detection of photons that lie close to
candidate signal $\pi^+$'s.  Following this calorimeter will be a muon veto
system in the form of an iron-scintillator hadronic calorimeter.  This
system is required to reject muons by $10^5:1$ for a pion acceptance
of 90\%.  Background photons traversing the beam hole in the forward
veto and muon vetoes will be detected in a ``hole'' veto system, and there
must also be an in-beam charged particle veto to catch $e^+$ and $e^-$
from photon conversions in the pion RICH.  The latter will be composed
of scintillating fiber planes and will also serve to tag beam
particles.

	Like E787 and E949, CKM features redundant background
rejection techniques.  The right hand graph in Fig.~\ref{fig:ckm}
shows the expected distribution of candidate events in missing mass
squared.  The shaded events are the signal, whereas the large peak is
due to $K^+ \to \pi^+ \pi^0$ background.

	This experiment is expected to start collecting data in
2007 or 2008.

\begin{figure}[ht]
 \includegraphics[angle=0, height=.4\textheight]{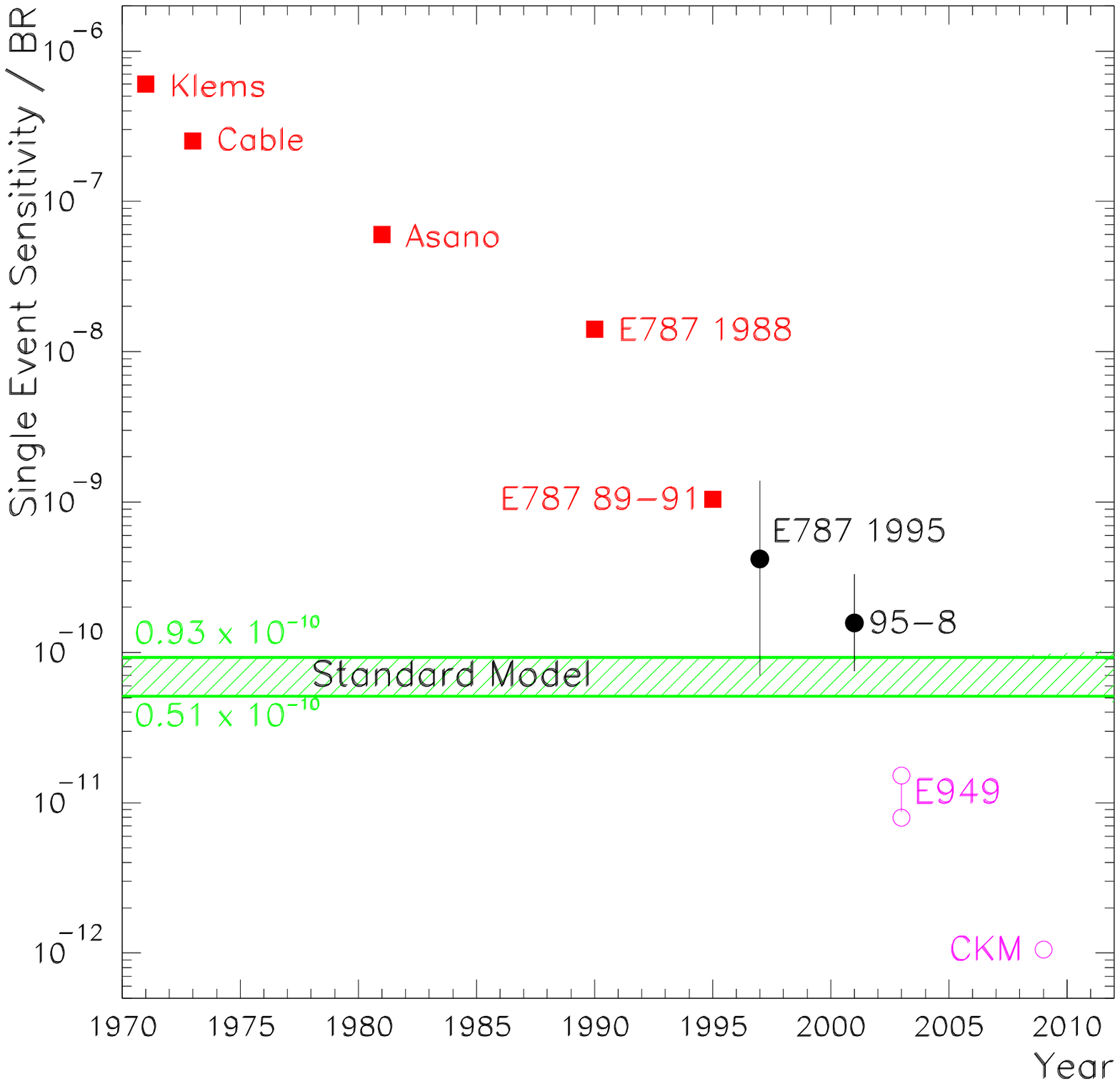}
  \caption{History and prospects for the study of \kpnnp.  Points without
error bars are single event sensitivities, those with error bars are 
measured branching ratio.
    \label{fig:prog} }
\end{figure}

Fig.~\ref{fig:prog} shows the history and expectations of progress in
studying \kpnnp.

\subsection{\kpnn0}
\kpnn0 is the most attractive target in the kaon system, since it is
direct CP-violating to a very good
approximation\cite{Littenberg:1989ix,Buchalla:1998ux} (\bkpnn0 $\propto
\eta^2$).  Like \kpnnp\ the
hadronic matrix element can be obtained from $K_{e3}$, but
unlike $K^+ \to \pi^+ \nu\bar\nu$,it has no significant contribution from charm.  As a result, the
intrinsic theoretical uncertainty connecting  \bkpnn0\ to
the fundamental SM parameters is only about 2\%.  
Note also that \bkpnn0~
is directly proportional to the square of $Im \lambda_t$ and
that $Im \lambda_t = - $$\cal{J}$$/[\lambda (1-\frac{\lambda^2}{2})]$
where $\cal{J}$ is the Jarlskog invariant\cite{Jarlskog:1985ht}.
Thus a measurement of \bkpnn0~determines the area of the unitarity
triangles with a precision twice as good as that on \bkpnn0~itself.

\begin{figure}[ht]
 \includegraphics[angle=0, height=.3\textheight]{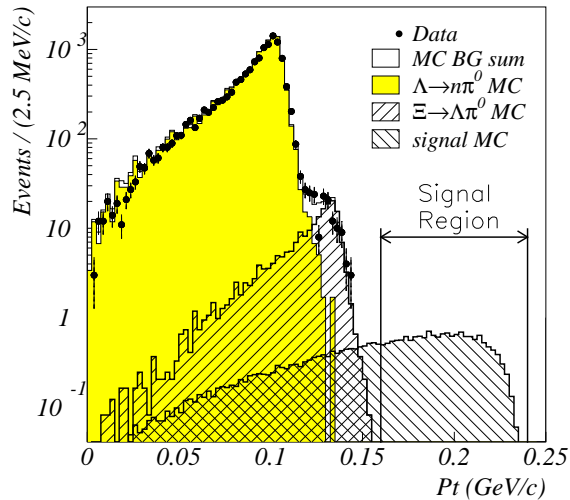}
  \caption{$P_t$ spectrum of $K_L \to \pi^0\nu\bar\nu$ candidates
from Ref.~\protect\cite{Alavi-Harati:1999hd}
    \label{fig:ktev_pnn0} }
\end{figure}

        \bkpnn0\ can be bounded indirectly by measurements of \bkpnnp\
through a nearly model-independent relationship pointed out by
Grossman and Nir\cite{Grossman:1997sk}.  The application of this to
the new E787 result yields \bkpnn0$<1.7 \times 10^{-9}$ at 90\% CL.
This is far tighter than the current direct experimental limit, $5.9
\times 10^{-7}$, obtained by KTeV\cite{Alavi-Harati:1999hd} using
$\pi^0 \to \gamma e^+ e^-$ decay.  To
actually measure \bkpnn0 at the SM level ($\sim 3 \times 10^{-11}$),
one will need to improve on this by some five orders of magnitude.
Fig.~\ref{fig:ktev_pnn0} shows the $p_T$ spectrum of events from that
experiment along with the calculated backgrounds.

The calculated background was 0.04 events.  This would give $\sim 800$ 
background events/SM signal event if just scaled up.  Clearly the
background rejection power of the experiment would need to be improved.
This might be possible, but the leading problem of this technique is
simply the factor $\sim 80$ loss of acceptance incurred by confining
oneself to Dalitz decays of the $\pi^0$.  To get anywhere near the
SM level, one will have to use the $\pi^0 \to \gamma\gamma$ branch and
this is in fact the method of all current and planned attempts to detect
\kpnn0.

\begin{figure}[ht]
 \includegraphics[angle=0, height=.3\textheight]{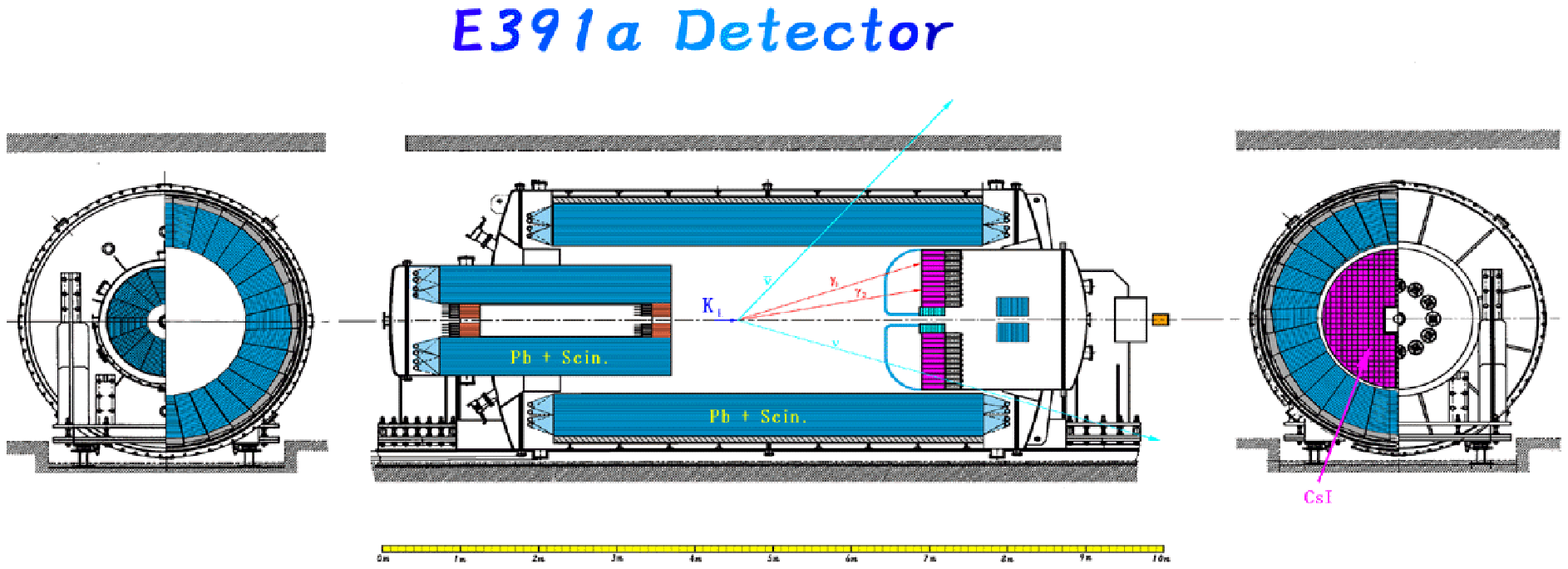}
  \caption{KEK E391a detector for $K_L \to \pi^0\nu\bar\nu$
    \label{fig:e391a} }
\end{figure}

The KEK E391a experiment\cite{Inagaki:1997gc} proposes to achieve a 
sensitivity of $\sim 3 \times 10^{-10}$/event which would better 
the indirect limit by a factor five, but would not fully bridge the gap 
between this limit and the SM level.
However, it will be sensitive to large BSM contributions to this decay
and will serve as a test for a future much more sensitive
experiment to be performed at the Japanese Hadron Facility.  E391a
features a carefully designed ``pencil'' beam with average momentum
$\sim 3.5$ GeV/c. Fig.~\ref{fig:e391a} shows a layout of the detector.

The photon veto system consists of two cylinders.  
The inner, more upstream barrel is intended to
suppress beam halo and reduce confusion from upstream $K_L$ decays.
Roughly 4\% of the $K_L$'s decay in the 2.4m fiducial region
between the end of the inner cylinder and the charged particle
veto in front of the photon detector.
Signal photons are detected in a multi-element CsI-pure crystal
calorimeter.  The entire apparatus will operate in
vacuum.  An advantage of this configuration is rather high
acceptance.  A particular challenge of this approach is to
achieve extremely low photon veto inefficiency.  Beamline construction
and tuning started in March 2000 and physics running is expected to
begin in Fall, 2003.

\begin{figure}[ht]
 \includegraphics[angle=0, height=.275\textheight]{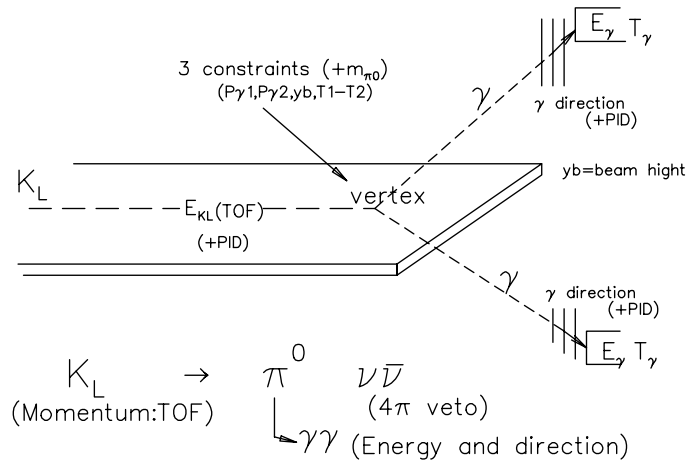}
  \caption{Principles of KOPIO $K_L \to \pi^0\nu\bar\nu$ experiment
    \label{fig:k_prin} }
\end{figure}

        The KOPIO experiment\cite{Bryman:2001hs} at BNL (E926) takes
a completely different approach, exploiting the intensity and flexibility of
the AGS to make a high-flux, low-energy, microbunched $K_L$ beam.
The principles of the experiment are illustrated in Fig.~\ref{fig:k_prin}.

The AGS proton beam will be microbunched at 25 MHz by imposing
upon it a train of empty RF buckets as it is extracted from the 
machine\cite{Glenn:1997dn}.  The neutral beam will be extracted at 
$\sim 45^{\rm o}$ to soften the $K_L$ spectrum
sufficiently to permit time-of-flight determination of the $K_L$ velocity.
The large production angle also softens the neutron spectrum 
so that they (and the $K_L$) are by and large below threshold for the
hadro-production of $\pi^0$'s.  The beam region will be evacuated to
$10^{-7}$ Torr to further minimize such production.  With a 10m beam channel 
and this low energy beam, the contribution of hyperons to the background will
be negligible.  The profile of the beam is ribbon-like to facilitate 
collimation of the large aperture and to provide an extra constraint for 
reconstruction of the decay vertex.  All possible quantities are measured:
in addition to the $K_L$ momentum, the photon angles as well as energies
and times.  In this way, powerful kinematic rejection of background is 
made possible.

\begin{figure}[ht]
 \includegraphics[angle=0, height=.44\textheight]{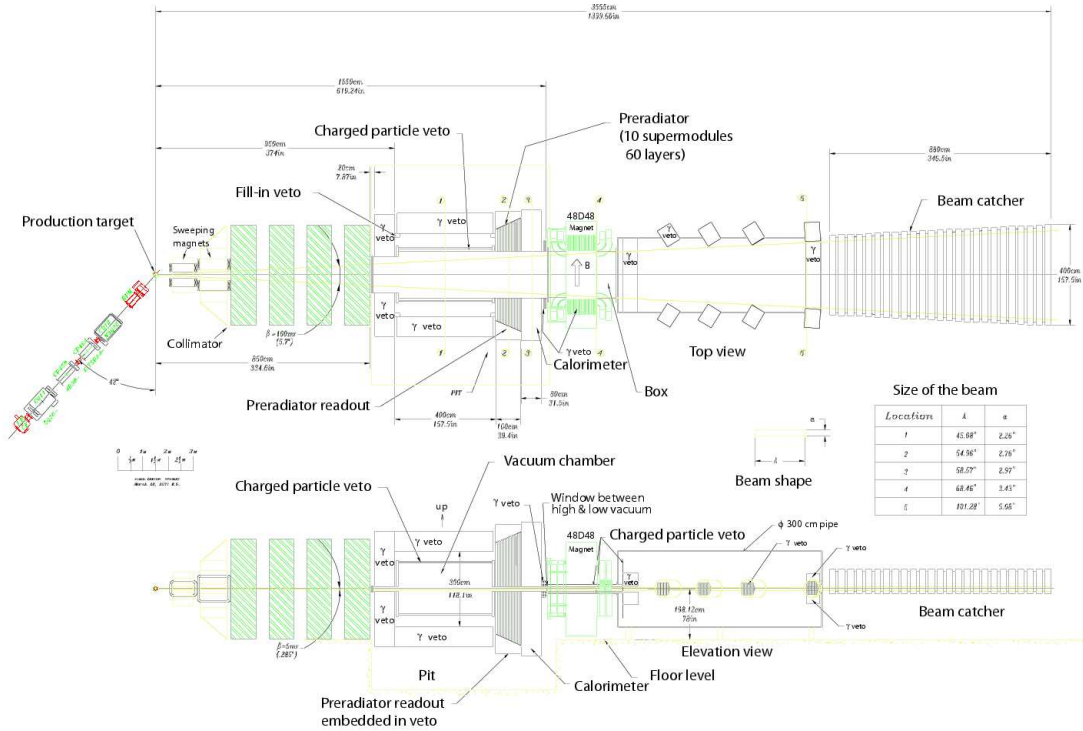}
  \caption{Layout of the KOPIO detector.
    \label{fig:kopio} }
\end{figure}

\begin{figure}[ht]
 \includegraphics[angle=0, height=.25\textheight]{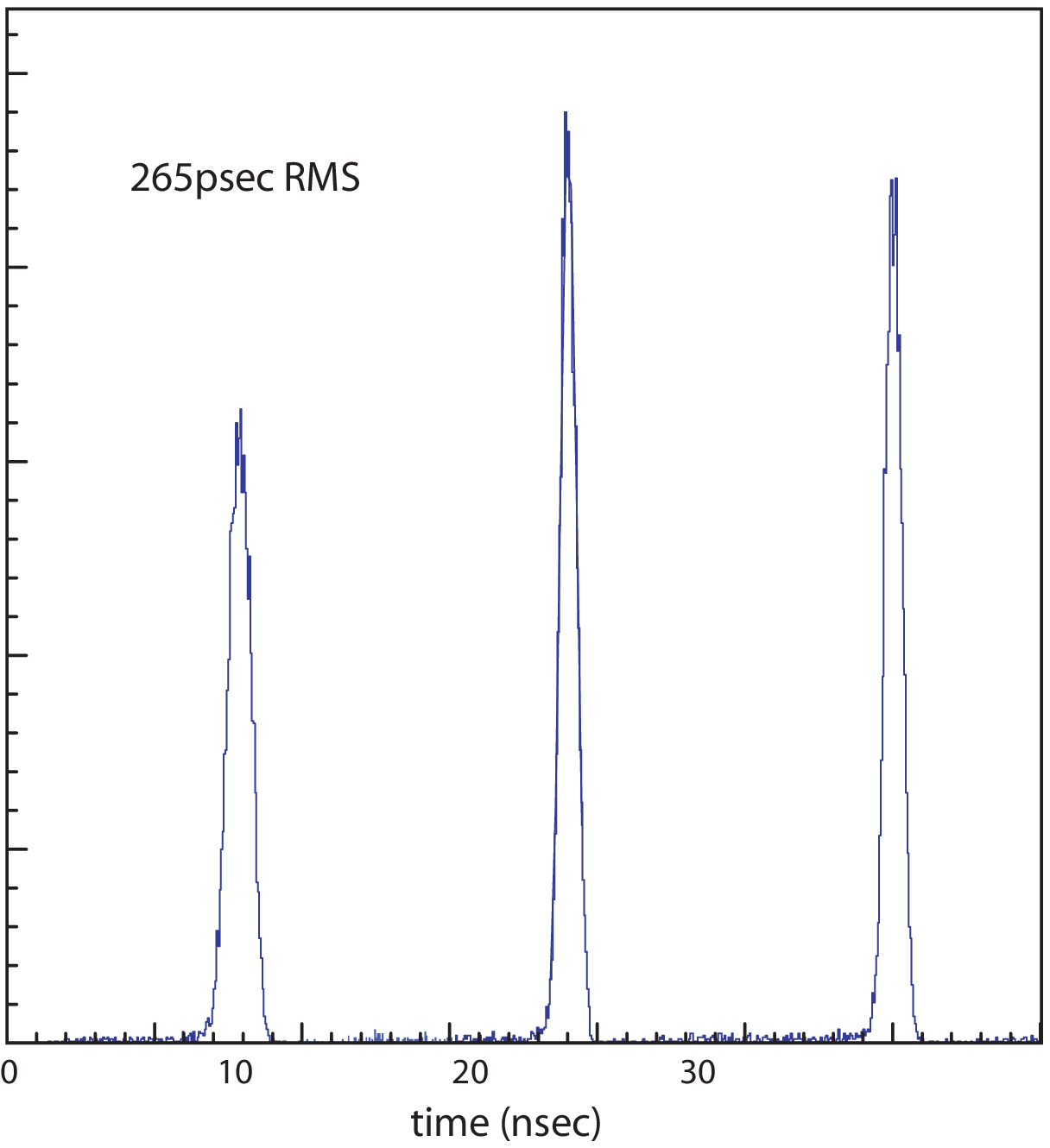}
 \includegraphics[angle=0, height=.25\textheight]{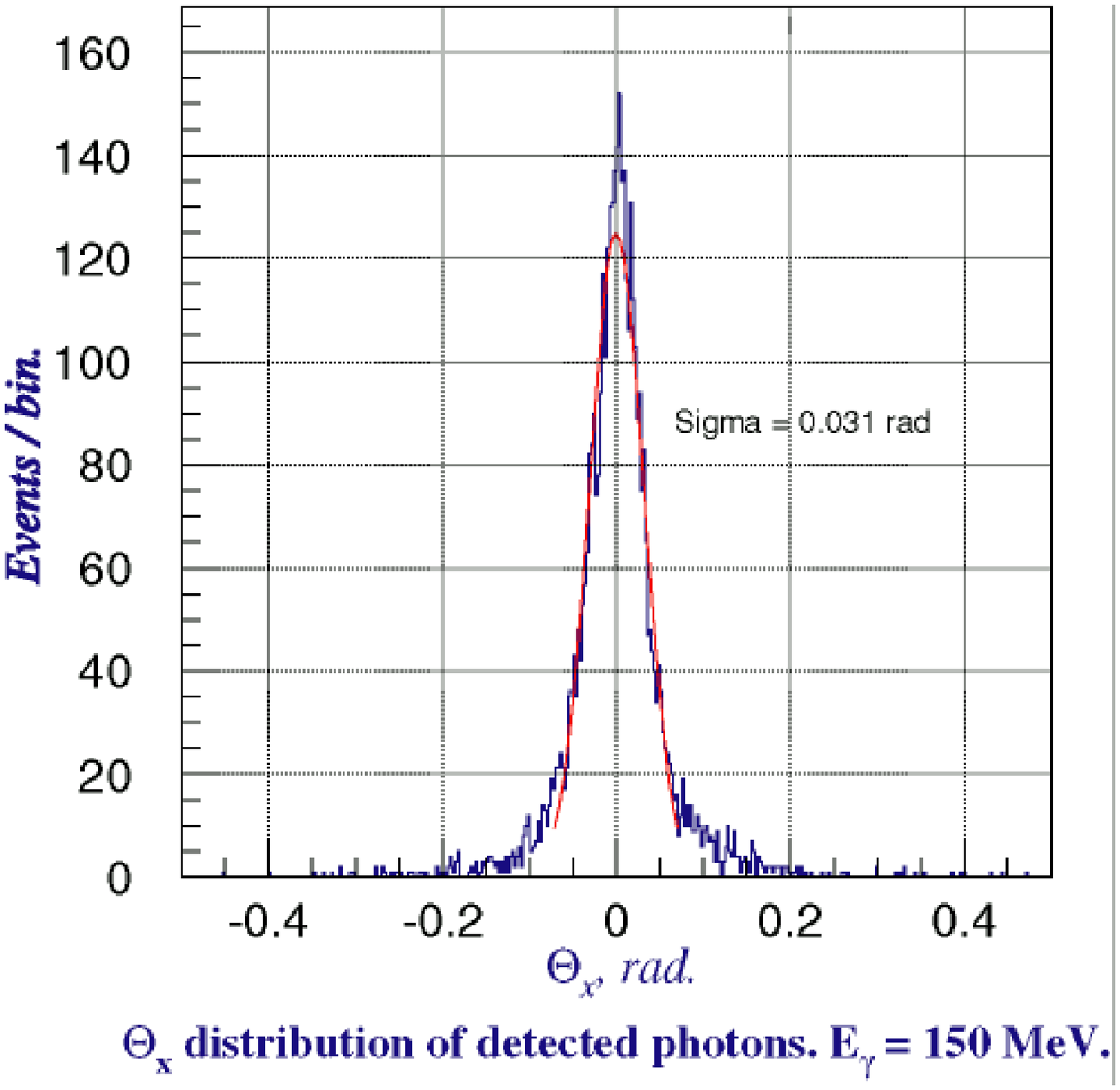}
  \caption{Tests of KOPIO components.  {\bf Left:} demonstration of microbunching
of the AGS proton beam.  {\bf Right:} angular resolution of prototype preradiator
in a tagged photon beam.
    \label{fig:k_tests} }
\end{figure}

The layout of the  experiment is shown in Fig.~\ref{fig:kopio}.  
$K_L$ decays from a $\sim 3$m fiducial region will be accepted.  Signal 
photons impinge on a 2 $X_0$ thick preradiator capable of measuring their 
direction to $\sim 30$mrad.  An alternating drift chamber/scintillator plane 
structure will also allow good measurement of the energy deposited in 
the preradiator.  A high-precision shashlyk calorimeter downstream of 
the preradiator will complete the energy measurement.  The photon
directional information will allow the decay vertex position to be determined.
This can be required to lie within the beam envelope, eliminating many
potentially dangerous sources of background.  
Combined with the target position and time of flight information,
the vertex information provides a measurement of the $K_L$ 3-momentum so that
kinematic constraints as well as photon vetoing are available to suppress
backgrounds.  The leading expected background is \kp0, which is 
initially some eight orders of magnitude larger than the predicted signal.  
However since $\pi^0$'s from this background have a unique energy in
the $K_L$ center of mass, a very effective kinematic cut can be
applied.  This reduces the burden on the photon veto system
surrounding the decay region to the point where the hermetic 
veto techniques proven in E787 are sufficient.  In fact most of
the techniques necessary for KOPIO have been proven in previous 
experiments or in prototype tests.  Fig.~\ref{fig:k_tests} shows results on
two of the more critical aspects of the experiment.

	On the left of the figure is the result of a recent test of
beam microbunching, showing an rms of 265 psec.  This is sufficient
for KOPIO's purposes, although it still needs to be demonstrated at a
25 MHz repetition rate.  On the right is a plot of photon angular
resolution obtained with a 6-plane prototype of the preradiator.  A
tagged beam at the National Synchrotron Light Source (NSLS) provided the
photons.  A resolution of $30 \,$mrad is observed for 150 MeV photons, in
line with GEANT simulation.  This resolution is sufficient for KOPIO.

	The electromagnetic calorimeter following the preradiator will
be a 5.2m $\times$ 5.2m array of high resolution shashlyk modules.  The
required resolution of 3-3.5\%/$\sqrt(E)$ has been demonstrated in prototypes
tested in the NSLS tagged photon beam.  

	The upstream and barrel vetoes will be 18 r.l. thick lead-scintillator
shower counters read out via wavelength-shifting fibers.  The demands on 
the performance of 
these counters are comparable to that demonstrated in the E787 barrel veto
which has similar structure.  Small prototypes have shown
good characteristics and full scale prototypes are currently in 
production.  It will also be necessary to veto within the beam, which is
very challenging but is facilitated by the low average energy of the
beam neutrons.  This will be accomplished by a series of lead-aerogel
shower counters (the ``catcher'' veto).  For the most part charged
particles created by the neutrons are below the Cerenkov threshold of
the aerogel and are so invisible to these counters.  

Another very important element is charged particle vetoing 
needed to eliminate backgrounds such as $K_L \to \pi^0 \pi^+ \pi^-$.
A very high performance system will be mounted in the decay region
vacuum and at the margins of the downstream beam pipe.  Behind the calorimeter
will be a dipole magnet with field oriented to sweep charged particles
traveling in the beam direction upwards or downwards into veto counters 
outside the beam profile.
	
KOPIO aims
to collect about $50$ \kpnn0\ events with a signal to background
ratio of 2:1.  This will permit $\eta$ to be determined to $\sim 10\%$,
given expected progress in measuring $m_t$ and $V_{cb}$.  KOPIO will
run during the $\sim$20 hours/day the AGS is not needed for injection 
into RHIC.  The experiment is presently in an R\&D phase.

\subsection{\kpll}

	The \kpll~are reactions initially thought to be more tractable 
experimentally than \kpnn0.  Like \kpnn0, in the SM they are sensitive to
$Im \lambda_t$, but in general they have different sensitivity to BSM
effects~\cite{Buras:1999da}.  Although their signatures are
intrinsically superior to that of \kpnn0, they are subject to a
serious background that has no analogue in the neutral lepton case: $K_L \to
\gamma\gamma\ell^+\ell^-$.  This process, a radiative correction to
$K_L \to \gamma\ell^+\ell^-$, occurs roughly $10^5$ times
more frequently than \kpll.  Kinematic cuts are
quite effective, but it is very difficult to improve the signal:background
beyond about $1:5$\cite{Greenlee:1990qy}.  Both varieties of  $K_L \to
\gamma\gamma\ell^+\ell^-$ have been observed, $B(K_L \to \gamma\gamma e^+ 
e^-)_{k_{\gamma}>5 MeV} = (5.84 \pm 0.15_{stat} \pm 0.32_{syst}) \times
10^{-7}$\cite{Alavi-Harati:2000tv} and $B(K_L \to \gamma\gamma \mu^+
\mu^-)_{m_{\gamma\gamma}> 1 MeV/c^2} = (10.4 {+7.5 \atop{-5.9}}_{stat} \pm 
0.7_{syst}) \times 10^{-9}$\cite{Alavi-Harati:2000hr}; both agree
reasonably well with theoretical prediction.   By comparison, the
branching ratio arising from SM short distance interactions,
$B^{direct}(K_L \to \pi^0 e^+ e^-)$, is predicted to be \cite{Buras:2001pn}
$(4.3 \pm 2.1)\times 10^{-12}$ and  $B^{direct}(K_L \to \pi^0 \mu^+ \mu^-)$
about five times smaller.

	In addition to this background, there are two other issues
that make the extraction of short-distance information from 
\kpll~rather challenging.  First, there is an indirect CP-violating
amplitude from the $K_1$ component of $K_L$ that is proportional to 
$\epsilon A(K_S \to \pi^0 e^+ e^-)$. 
It is of the same order of magnitude as the direct CP-violating 
amplitude and interferes with it, yielding~\cite{D'Ambrosio:1998yj}:
\be
B(K_L \to \pi^0 ee)_{CPV} \approx \left [15.3 a_S^2 - 6.8 a_s \frac{Im \lambda_t}{10^{-4}} + 2.8 \left(\frac{Im \lambda_t}{10^{-4}}\right)^2\right] \times 10^{-12}
\label{cpks}
\ee
where 
\be
B(K_S \to \pi^0 ee) \approx 5.2 a_S^2 \times 10^{-9}
\label{ksbr}
\ee
One can get a very rough estimate of $a_S$ from the related process
$K^+ \to \pi^+ e^+ e^-$, on which there is now rather good 
data. The corresponding parameter is measured in that decay to be  
$a_{+} = -0.587 \pm 0.010$~\cite{Appel:1999yq}.  However the dangers
of extrapolating from $K^+ \to \pi^+ e^+ e^-$ to $K_S \to \pi^0 e^+ e^-$
have been pointed out in Ref.~\cite{Littenberg:1993qv}. Thus the size of 
the indirect CP-violating contribution will be predictable if and when 
$B(K_S \to \pi^0 e^+ e^-)$ is measured, hopefully by the upcoming NA48/1
experiment~\cite{NA48-1}.  At the moment our information is limited to
$|a_S|< 5.2$ from the NA48 result $B(K_S \to \pi^0 ee) < 1.4 
\times 10^{-7}$ at 90\% CL~\cite{Lai:2001jf}.

\begin{figure}[ht]
 \includegraphics[angle=0, height=.28\textheight]{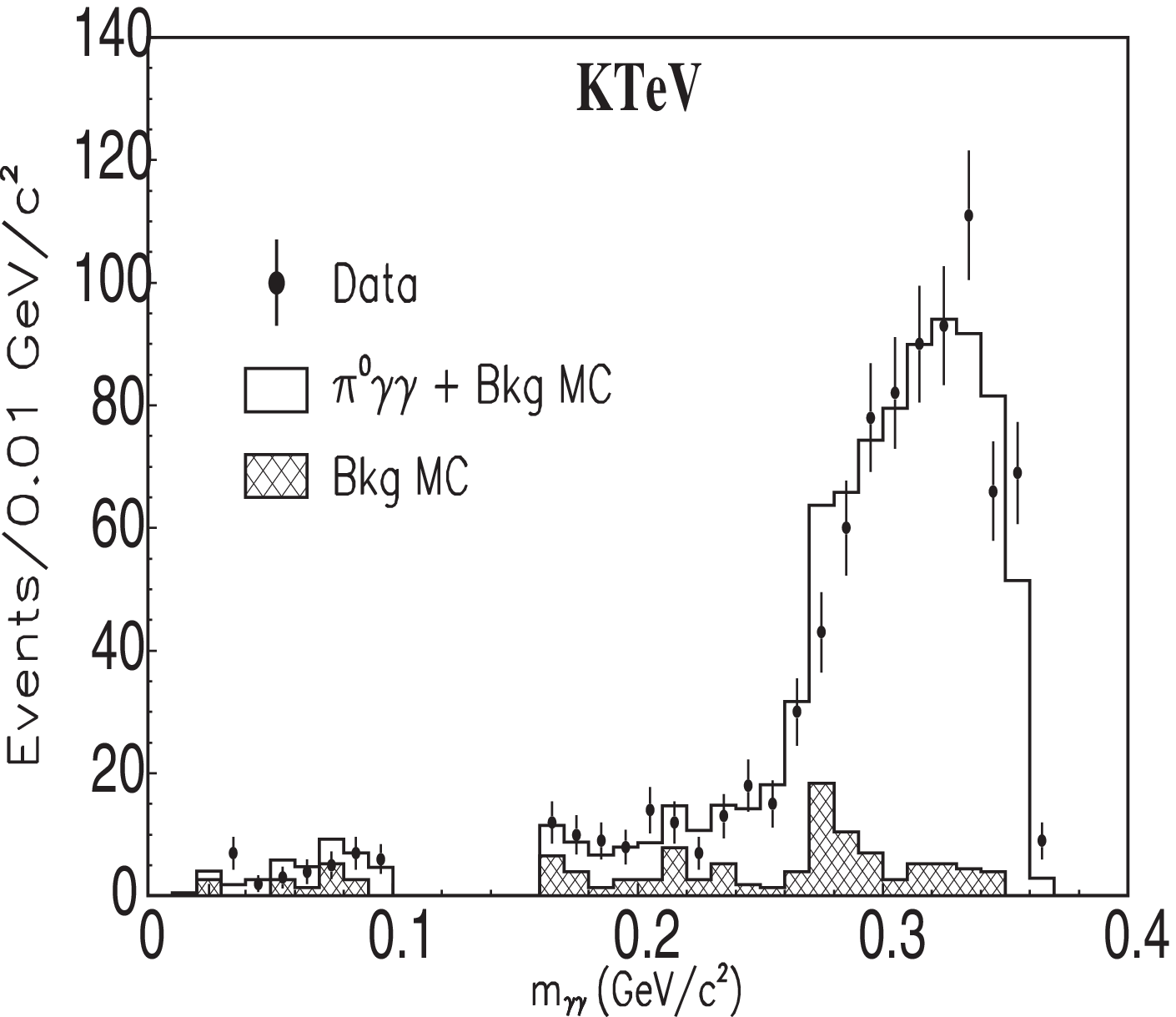}
 \includegraphics[angle=0, height=.27\textheight]{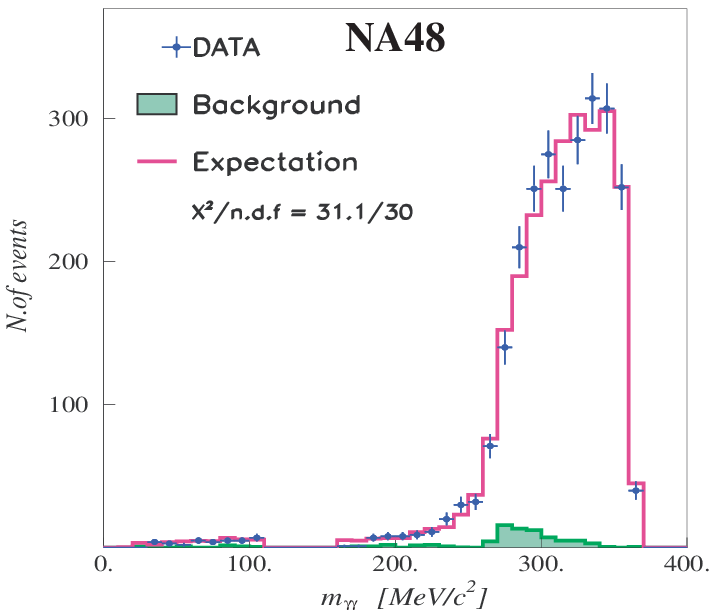}
  \caption{$m_{\gamma\gamma}$ spectrum of $K_L \to \pi^0 \gamma\gamma$
candidates from (left) KTeV~\protect\cite{Alavi-Harati:1999mu} and (right) 
NA48~\cite{Lai:2002kf}.  The striking threshold at $m_{\gamma\gamma}
\approx 2 m_{\pi}$ is due to the amplitude $A(y,z)$ in $\chi$PT (or in
pion loop models~\protect\cite{Sehgal:1972hs}).
    \label{fig:pgg0} }
\end{figure}

Another contribution of similar order, mediated by $K_L \to \pi^0
\gamma\gamma$, is {\it CP-conserving}.  In principle
this contribution can be predicted from measurements of the branching
ratio and kinematic distributions of $K_L \to \pi^0 \gamma\gamma$, and
thousands of these events have been observed.  
The matrix element for this decay is given by~\cite{Ecker:1988hd}:
\bea
{\mathcal M}(K_L \to \pi^0 \gamma \gamma) = \frac{G_8 \alpha}{4 \pi}
\epsilon_{\mu}(k_1) \epsilon_{\nu}(k_2)  \big [ A(k^{\mu}_2 k^{\nu}_1 - 
k_1 \cdot k_2 g^{\mu \nu}) + \cr
 B \frac{2}{m^2_K}(p_k \cdot k_1 k^{\mu}_2 p_K^{\nu}
+p_K \cdot k_2 k^{\nu}_1 p_K^{\mu} -k_1 \cdot k_2 p_K^{\mu} p^{\nu}_K - g^{\mu\nu}
p_K \cdot k_1 p_K \cdot k_2) \big ]
\label{pggm}
\eea
where $k_1$ and $k_2$ refer to the photons. 
$A$ and $B$ refer to the $J_{\gamma\gamma} = 0$ and
$J_{\gamma\gamma} = 2$ amplitudes respectively, and $G_8$ is
the octet coupling constant in $\chi$PT.
Eqn.~\ref{pggm} leads to 
\be 
\frac{\partial^2 \Gamma(K_L \to \pi^0 \gamma\gamma)}{\partial y \partial z} = \frac{m_K}{2^9 \pi^3}
\left [ z^2 |A + B|^2 + (y^2- \frac{1}{4}\lambda(1,r^2_{\pi},z))^2 |B|^2 \right ]
\label{pggm2}
\ee
where $z \equiv (k_1 + k_2)^2/m^2_K$, $y \equiv p_K \cdot (k_1 - k_2)/m^2_K$,
$r_{\pi} \equiv m_{\pi}/m_K$ and $\lambda(a,b,c) \equiv a^2 + b^2 + c^2
-2(ab+ac+bc)$.  Since the effect of $A$
on $K_L \to \pi^0 e^+ e^-$ is greatly suppressed by helicity conservation
and $B = 0$ at leading order in $\chi$PT, it was initially thought
that the CP-conserving contribution to $K_L \to \pi^0 e^+ e^-$ would
be very small\footnote{There is very little helicity suppression for 
$K_L \to \pi^0 \mu^+ \mu^-$
so that the CP-conserving branching ratio is relatively larger.}.  
However the possibility of a substantial vector meson
dominance (VDM) contribution to $B$ was pointed out by 
Sehgal~\cite{Sehgal:1988ej}. Such a contribution can arise at 
$\cal{O}$$(p^6)$ in $\chi$PT.  Indeed, early measurements
of $B(K_L \to \pi^0 \gamma\gamma)$~\cite{Barr:1990hc,Papadimitriou:1991iw} 
showed that although the simple $\cal{O}$$(p^4)$ calculation was in reasonable
agreement with the $m_{\gamma\gamma}$ spectrum, it underestimated the decay 
rate 
by a factor $\sim 3$.  There has since been a good deal of theoretical work 
devoted to remediating this~\cite{Cohen:1993ta,Cappiello:1993kk,
Heiliger:1993uh,Kambor:1994tv,D'Ambrosio:1997sw}.  
Although a full $\mathcal{O}(p^6)$
calculation is not possible at present, in this work the $\mathcal{O}(p^4)$ 
calculation was  improved by ``unitarity corrections'' and the addition of a 
VDM contribution characterized by a single parameter $a_V$.  This produced
satisfactory agreement with the observed branching ratio, at least until
the recent, more precise measurements.  A similar 
approach~\cite{D'Ambrosio:1996zx} was successful in predicting the 
characteristics of the closely related decay $K^+ \to \pi^+ \gamma\gamma$ 
that was measured by AGS E787\cite{Kitching:1997zj}.  The recent data 
on $K_L \to \pi^0 \gamma\gamma$ is 
summarized in Table~\ref{kpgg} and the $m_{\gamma\gamma}$ spectra are shown in
Fig.~\ref{fig:pgg0}.  Unfortunately the two newest results, from
KTeV~\cite{Alavi-Harati:1999mu} and from NA48~\cite{Lai:2002kf} disagree by 
nearly $3 \sigma$ in branching ratio.  Their spectra also differ rather
significantly, leading to 
differing extracted values of $a_V$ and thus to disagreement in 
their predictions for $B^{CP-cons}(K_L \to \pi^0 e^+ e^-)$,
as seen in Table~\ref{kpees}.  This is inimical to
the prospects of measuring $B^{direct}(K_L \to \pi^0 e^+ e^-)$.

\begin{table}[htbp]
\begin{tabular}{lll} \hline
Exp/Ref & $B(K_L \to \pi^0 \gamma \gamma ) \cdot 10^6$ & $a_V$ \\
\hline
KTeV~\cite{Alavi-Harati:1999mu}  & $1.68 \pm 0.07_{stat} \pm 0.08_{syst} $                & $-0.72 \pm 0.05 \pm 0.06 $              \\
NA48~\cite{Lai:2001jf}           & $1.36 \pm 0.03_{stat} \pm 0.03_{syst} \pm 0.03_{norm}$ & $-0.46 \pm 0.03 \pm 0.03 \pm 0.02_{theor} $\\
\end{tabular}
\caption{\it Recent results on \kpgg.}
\label{kpgg}
\end{table}

Moreover the use of this formalism to predict $B^{CP-cons}(K_L \to
\pi^0 e^+ e^-)$ from \kpgg~ has recently been reexamined
by Gabbiani and Valencia~\cite{Gabbiani:2001zn}. They point out
that the use of a single parameter $a_V$ artificially correlates the 
$A$ and $B$ amplitudes.  They show that a three parameter
expression inspired by $\mathcal{O}(p^6)$ $\chi$PT fits the KTeV data
quite as well as the conventional one based on $a_V$ and gives a
significantly different prediction for $B^{CP-cons}(K_L \to
\pi^0 e^+ e^-)$ as seen in Table~\ref{kpees}.  In a subsequent paper
~\cite{Gabbiani:2002bk}, they find they can make a good
simultaneous fit to the NA48 decay rate and spectrum using the same
technique and point out that this is not possible using just $a_V$.  The
predictions of their three parameter fits for $ B^{CP-cons}(K_L \to
\pi^0 e^+ e^-)$ also differ markedly between the two experiments as
shown in Table~\ref{kpees}.

	Finally there are significant uncertainties in the extraction
of the dispersive contribution to
$ B^{CP-cons}(K_L \to \pi^0 e^+ e^-)$~\cite{Donoghue:1995yt, Gabbiani:2001},
which is similar in size to the absorptive contribution and so not at
all negligible.

\begin{table}[htbp]
\begin{tabular}{cccc} \hline
Exp. &  $a_V$ fit from  &  $a_V$ fit by &  3 parameter fit by \\
     & experimental paper  &  Gabbiani \& Valencia &  Gabbiani \& Valencia\\
\hline
KTeV  &  $(1.0-2.0) \cdot 10^{-12}$  &$4.8 \cdot 10^{-12}$ &  $7.3 \cdot 10^{-12}$ \\
NA48  &  $(0.47 {+0.22 \atop -0.18}) \cdot 10^{-12}$ & $(1.38 {+0.09 \atop -0.21}) \cdot 10^{-12}$ &$(0.46 {+0.22 \atop -0.17}) \cdot 10^{-12}$  \\
\end{tabular}
\caption{\it Predictions for $B^{CP-cons}(K_L \to \pi^0 e^+ e^-)$.}
\label{kpees}
\end{table}

Thus both the theoretical and experimental situations are quite
unsettled at the moment.  Depending on which $K_L \to \pi^0
\gamma\gamma$ data one uses and how one uses it, values from $0.25
\times 10^{-12} $ to $7.3 \times 10^{-12}$ are predicted for
$B^{CP-cons}(K_L \to \pi^0 e^+ e^-)$.

\begin{figure}[ht]
 \includegraphics[angle=0, height=.30\textheight]{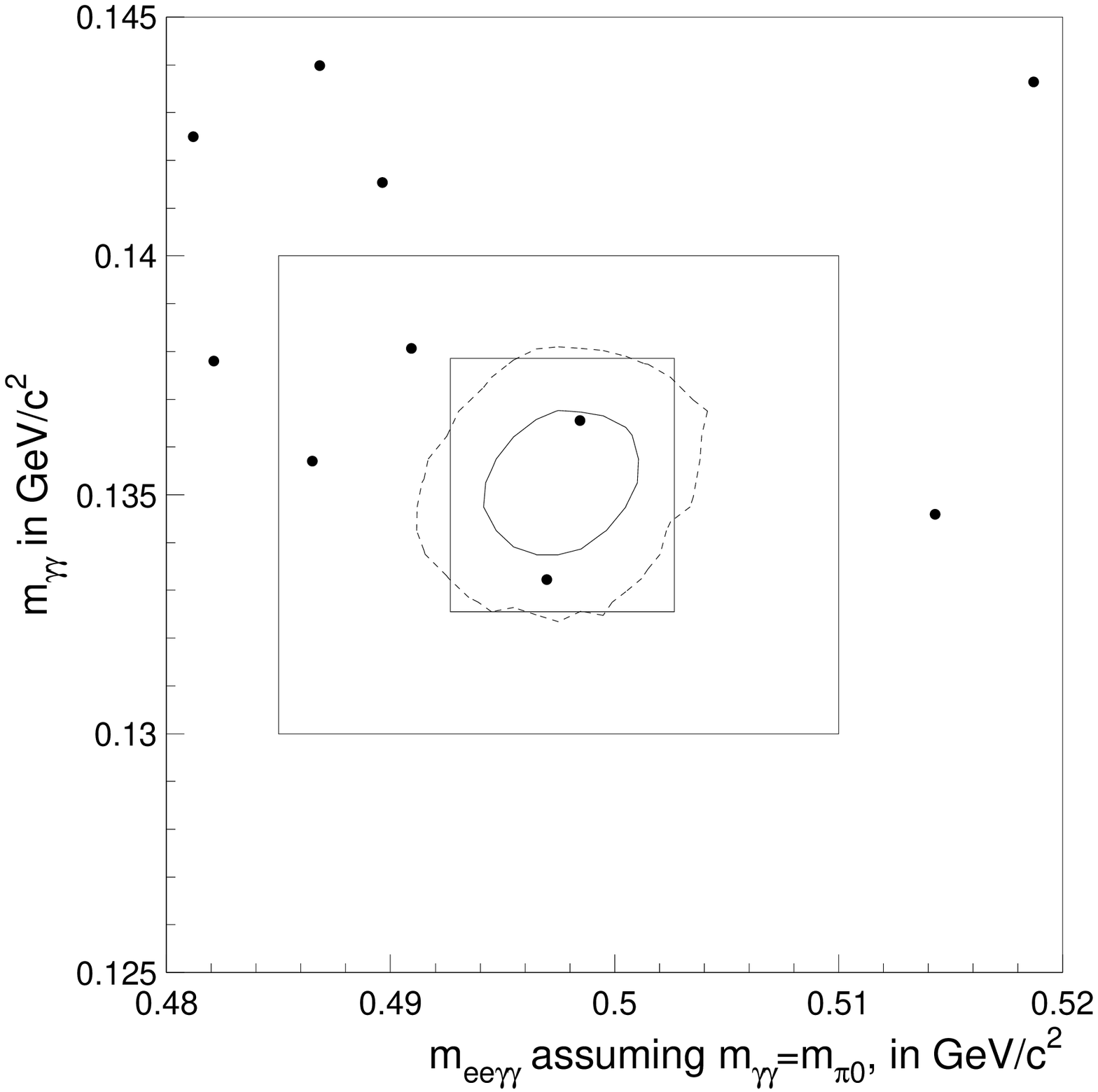}
 \includegraphics[angle=0, height=.325\textheight]{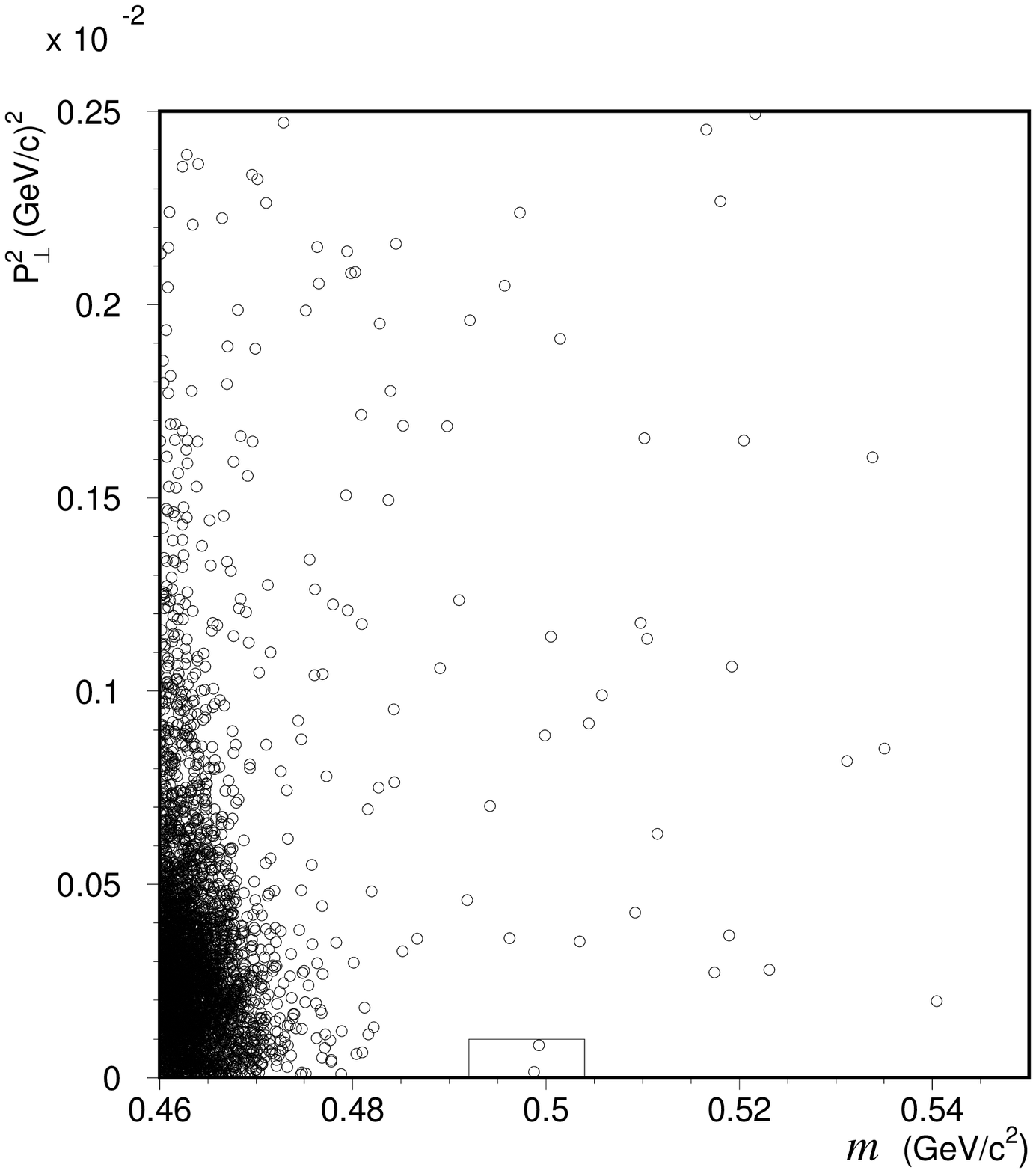}
  \caption{Signal planes showing candidates for $K_L \to \pi^0 e^+ e^-$ (left
from Ref.~\protect\cite{Alavi-Harati:2000sk}) and $K_L \to \pi^0 \mu^+ \mu^-$ (right
from Ref.~\protect\cite{Alavi-Harati:2000hs}).
    \label{fig:llpi0} }
\end{figure}

	The current experimental status of \kpll~is summarized in Table
~\ref{Kpll} and Fig.\ref{fig:llpi0}.  A factor $\sim 2.5$ more data is 
expected from the KTeV
1999 run, but as can be seen from the table and figure, background is already
starting to be observed at a sensitivity roughly 100 times short of
the expected signal level.  

\begin{table}[h]
\begin{tabular}{lllll} \hline
Mode & 90\% CL upper limit & Est. bkgnd.& Obs. evts. &Ref. \\
\hline
$K_L \to \pi^0 e^+ e^-$     & $5.1 \times 10^{-10}$ & $1.06\pm 0.41$ & 2 & \cite{Alavi-Harati:2000sk} \\
$K_L \to \pi^0 \mu^+ \mu^-$ & $3.8 \times 10^{-10}$ & $0.87\pm 0.15$ & 2 & \cite{Alavi-Harati:2000hs} \\
\end{tabular}
\caption{\it Results on \kpll.}
\label{Kpll}
\end{table}

\begin{figure}[ht]
 \includegraphics[angle=90, height=.25\textheight]{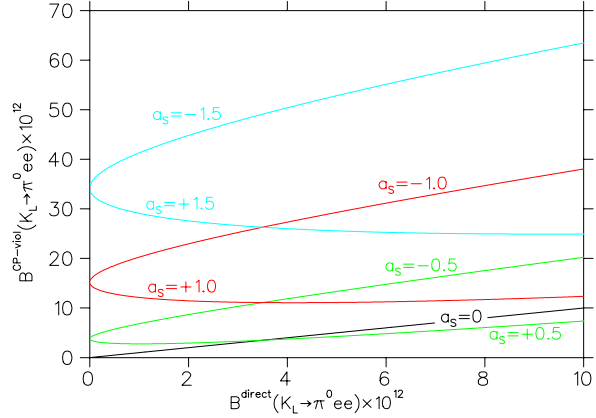}
  \caption{Relationship between $B^{CP-viol}(K_L \to \pi^0 e^+ e^-)$ and
$B^{direct}(K_L \to \pi^0 e^+ e^-)$ for various values of $a_S$.
    \label{fig:kscp} }
\end{figure}

One can get a feeling for the implications of the current data by
calculating what it would take to get a 20\% measurement of
$B^{direct}(K_L \to \pi^0 e^+ e^-)$ at the SM-predicted level, 
{\it i.e.} with precision
comparable to what is being discussed for $K_L \to \pi^0 \nu\bar\nu$.
The most straightforward case is that where the state-mixing and
CP-conserving components turn out to be negligible.  One then 
extrapolates the $K_L \to \gamma\gamma e^+ e^-$ background in the
current experiment, $0.91 \pm 0.26$\cite{Alavi-Harati:2000sk}, assuming that 
it gets statistically better determined $\propto$ 1/sensitivity.  One finds 
that without improvements in the background discrimination
power, it would require an experiment with a single event sensitivity
of $0.77 \times 10^{-14}$, {\it i.e.} about 13,000 times that of the
present one.  Although not on the near horizon, sensitivities in this range 
have been discussed in connection 
with a proposed high-intensity proton driver at CERN~\cite{Belyaev:2001kz}.
The result is rather insensitive to the presence of the CP-conserving term, 
{\it but only as long as it is very well determined}, which, as
discussed above, is not presently the case.  Any uncertainty on the level 
to be subtracted  dilutes the sensitivity and, for a given fractional 
uncertainty,  the larger the CP-conserving 
component, the larger the impact on the sensitivity.
For example, if $B^{CP-cons}(K_L \to \pi^0 e^+ e^-)$ is
known to be $2 \times 10^{-12}$ to 25\% precision, it degrades the
precision on $B^{direct}(K_L \to \pi^0 e^+ e^-)$ by only 10\%.  However,
if $B^{CP-cons}(K_L \to \pi^0 e^+ e^-)$ is known to be $4.3 \times
10^{-12}$ to 25\%, this degrades the precision on $B^{direct}(K_L \to
\pi^0 e^+ e^-)$ by 40\%.  The effects of the indirect CP-violating
contribution are still more problematical because of the interference. 
Using Eqn.~\ref{cpks}, Fig.~\ref{fig:kscp} shows the measured CP-violating
branching ratio vs the direct branching ratio for various plausible 
values of $a_S$.  It is clear that for certain cases, such as $a_S = 1$,
there is very little sensitivity to the direct branching ratio.  On
the other hand, since the CP-violating branching ratio can be much
enhanced by the indirect part, for some values of $a_S$, the measurement
gets considerably easier. Take the case of a negligible CP-conserving 
component and $a_S = -1$ and further assume that we have very good knowledge 
of $B(K_S \to \pi^0 e^+ e^-)$ and therefore of $|a_S|$.  Then if 
$B^{direct}(K_L \to \pi^0 e^+ e^-) = 4.3 \times 10^{-12}$, 
$B^{CP-viol}(K_L \to \pi^0 e^+ e^-) = 27.9 \times 10^{-12}$.  The 
expected result of the experiment 
described above would be 15,460 events over a background of 11,830,
{\it i.e.} we'd have  $B^{CP-viol}(K_L \to \pi^0 e^+ e^-) = 
(27.9 \pm 1.24)\times 10^{-12}$.   This high value would immediately
determine a negative sign for $a_S$ and the result would then
yield $\pm$15\% errors on 
$B^{direct}(K_L \to \pi^0 e^+ e^-)$, which more than meets our $\leq$ 20\%
criterion.  However, one will not know going in how well the experiment
will work, since the sign of $a_S$ won't be known in advance.
\footnote{It should also be kept in mind that Eqns.~\ref{cpks} and~\ref{ksbr} 
are approximations.}

\begin{figure}[h]
 \includegraphics[angle=0, height=.325\textheight]{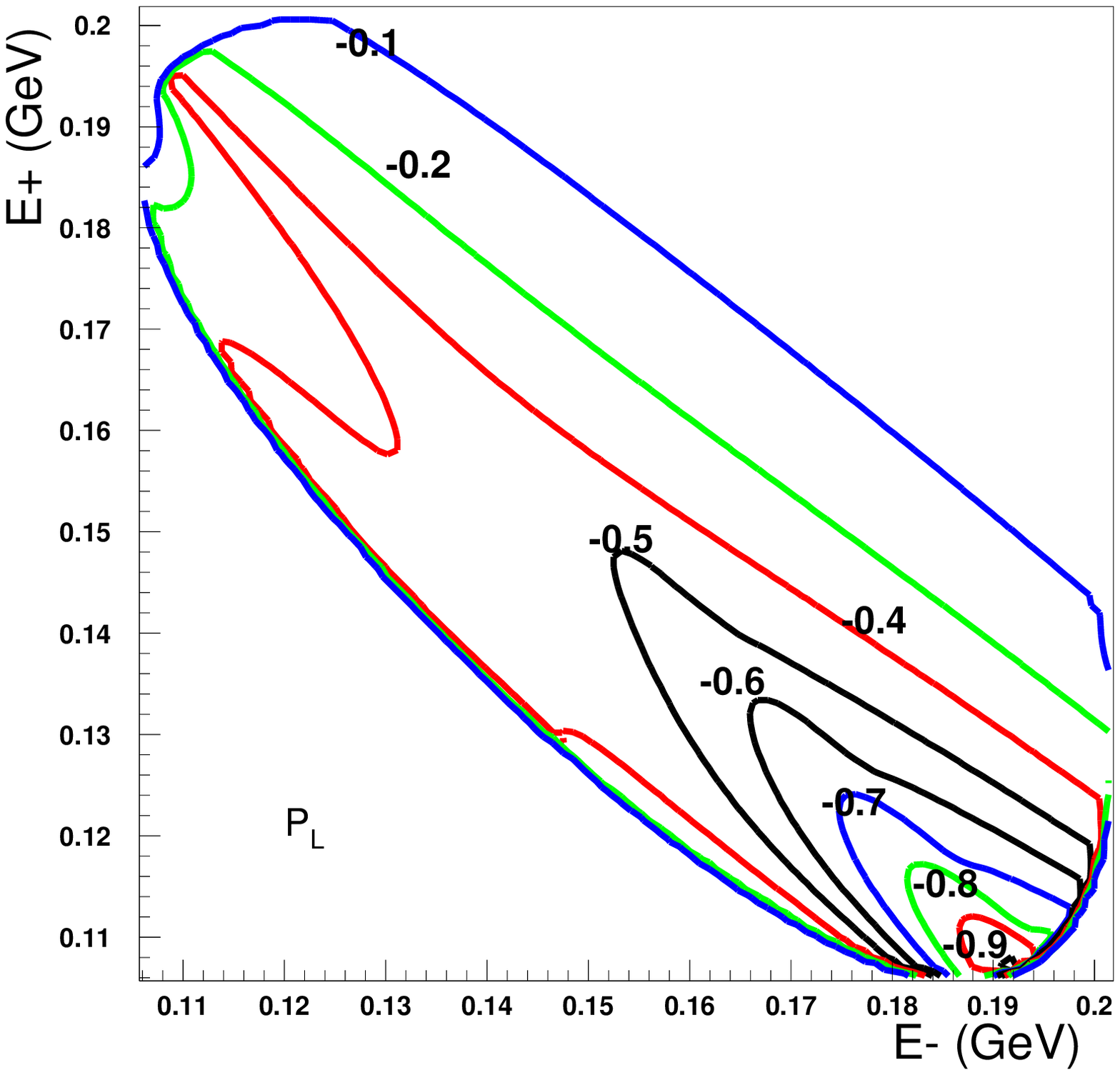}
\includegraphics[angle=0, height=.325\textheight]{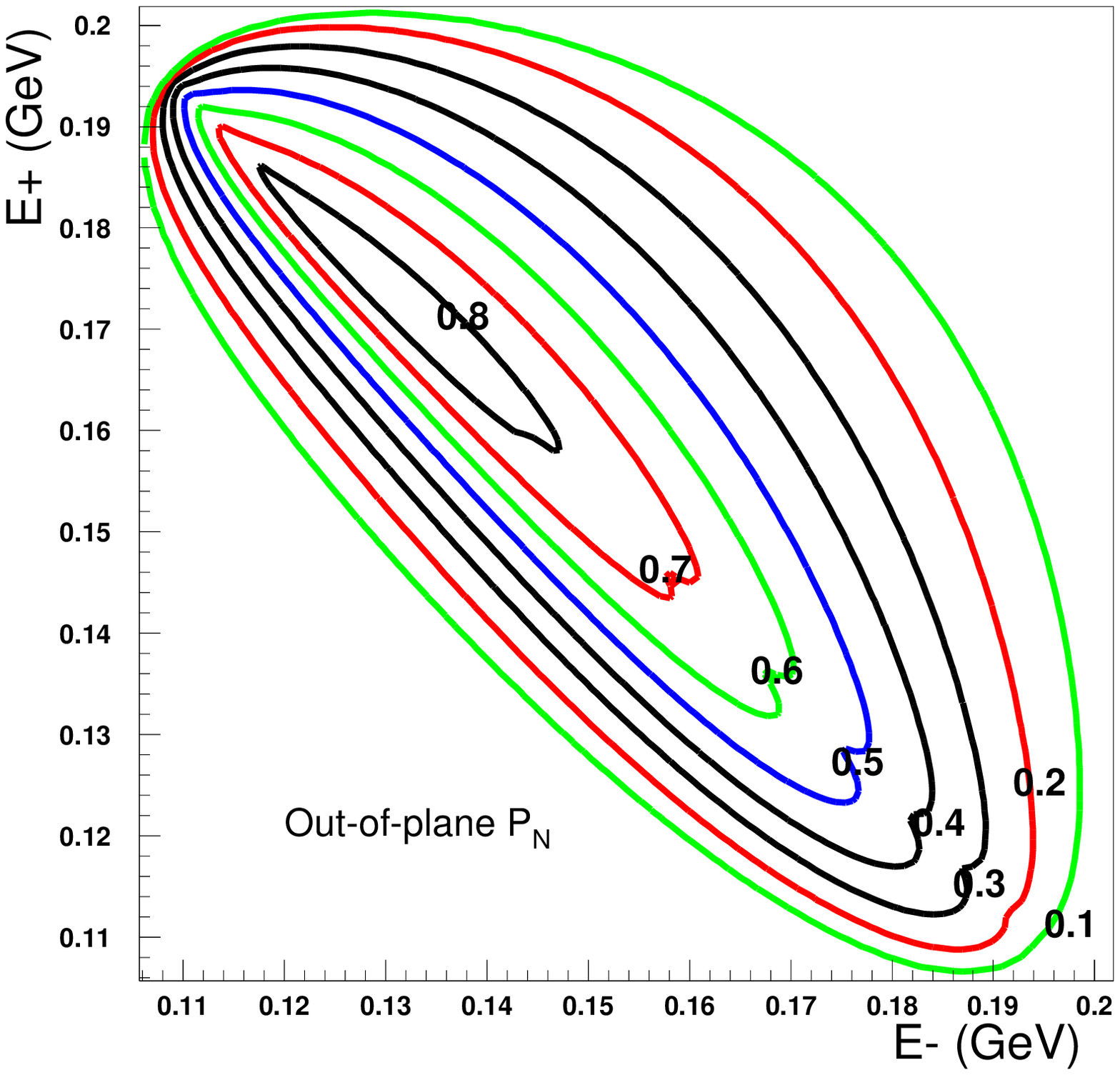}
  \caption{$\mu^+$ polarizations in \kpmm from Ref~\protect\cite{Diwan:2001sg} 
plotted against the muon cm energies.  {\bf Left:} longitudinal polarization.
{\bf Right:} out-of-plane polarization.  
    \label{fig:md} }
\end{figure}

	Thus, to make a useful measurement will require a 4 order
of magnitude increase in signal statistics and both theoretical and
experimental advances in the ancillary modes \kpgg~and $K_S \to \pi^0
e^+ e^-$, and still might not succeed.  Various approaches for
mitigating these problems have been suggested over the years including
studies of the Dalitz Plot \cite{Donoghue:1987aw,Donoghue:1995yt}, 
the $\ell^+ - \ell^-$ energy asymmetry ~\cite{Sehgal:1988ej}
\cite{Donoghue:1995yt}, the time development~\cite{Kohler:1995rb,
Belyaev:2001kz,Donoghue:1995yt}, or all three~
\cite{Littenberg:1988cy}.  An innovative approach has recently
been suggested \cite{Diwan:2001sg} in which muon polarization in $K_L
\to \pi^0 \mu^+ \mu^-$ as well as kinematic distributions are
exploited.  It's been known for many years that the $\mu^+$ out-of-plane
transverse polarization in this decay is sensitive to both the direct and
indirect CP-violating amplitudes~\cite{Ecker:1988hd}, and that one
might be able to determine the sign of $a_S$ through this.  However,
the individual effects of the direct and indirect amplitude are not
easy to untangle using the transverse polarization alone.
Ref.~\cite{Diwan:2001sg} assesses the potential of a number of polarization 
observables, and points out that the P-odd $\mu^+$ longitudinal
polarization is proportional to the direct CP-violating amplitude
alone, even though it is not in itself a CP-violating quantity,
whereas the branching ratio, the energy asymmetry and the
out-of-plane polarization depend on both indirect and direct CP-
violating amplitudes.  As shown in Fig.~\ref{fig:md}, the
polarizations involved turn out to be extremely large so that 
even with the polarization-diluting effect of the $K_L \to \gamma\gamma 
\mu^+ \mu^-$ background, enormous numbers of events may not be required to 
extract the direct amplitude.  This method is reasonably clean theoretically
and can determine the sign as well as the magnitude of that amplitude.

\section{Pion beta decay}

The sole current dedicated rare pion decay experiment is the PIBETA
experiment at PSI. The primary objective of PIBETA is a precision
measurement of the decay $\pi^+ \to \pi^0 e^+ \nu$.  This is an
example of a decay suppressed only by kinematics to the $10^{-8}$
level.  The main interest in this decay is the determination of the
CKM matrix element $V_{ud}$.  There's a long-standing mystery in
the experimental verification of the unitarity of the CKM matrix:
the sum of the squares of the moduli of the first row of the CKM
matrix does not quite add up to 1.  Using the latest PDB
values~\cite{Hagiwara:2002pw}, $|V_{ud}| = 0.9734 \pm 0.0008$,
$|V_{us}| = 0.2196 \pm 0.0026$, $|V_{ub}| = 0.0036 \pm 0.0007$ yields:
\be
|V_{ud}|^2 + |V_{us}|^2 + |V_{ub}|^2 = 0.9957 \pm 0.0019
\label{unitary1}
\ee
{\it i.e.} a 2.2$\,\sigma$ effect.  The above value of $|V_{ud}|$ comes from 
measurements of nuclear beta decays and from neutron decay\footnote{Depending 
on how various data are weighted and what theoretical input is used, this 
discrepancy can be made as large as 4$\,\sigma$.}.  To explain
the discrepancy of Eqn.~\ref{unitary1}, $|V_{ud}|$ would have to increase
by 0.22\%.  This would result in a 0.45\% increase in $B(\pi^+ \to \pi^0
e^+ \nu)$.  There's $\leq$0.1\% theoretical uncertainty in the connection
between this branching ratio and $|V_{ud}|$~\cite{Marciano:1986pd},
so it would be of considerable
interest if a measurement on the $\leq$0.5\% level could be made.

\begin{figure}[h]
 \includegraphics[angle=0, height=.25\textheight]{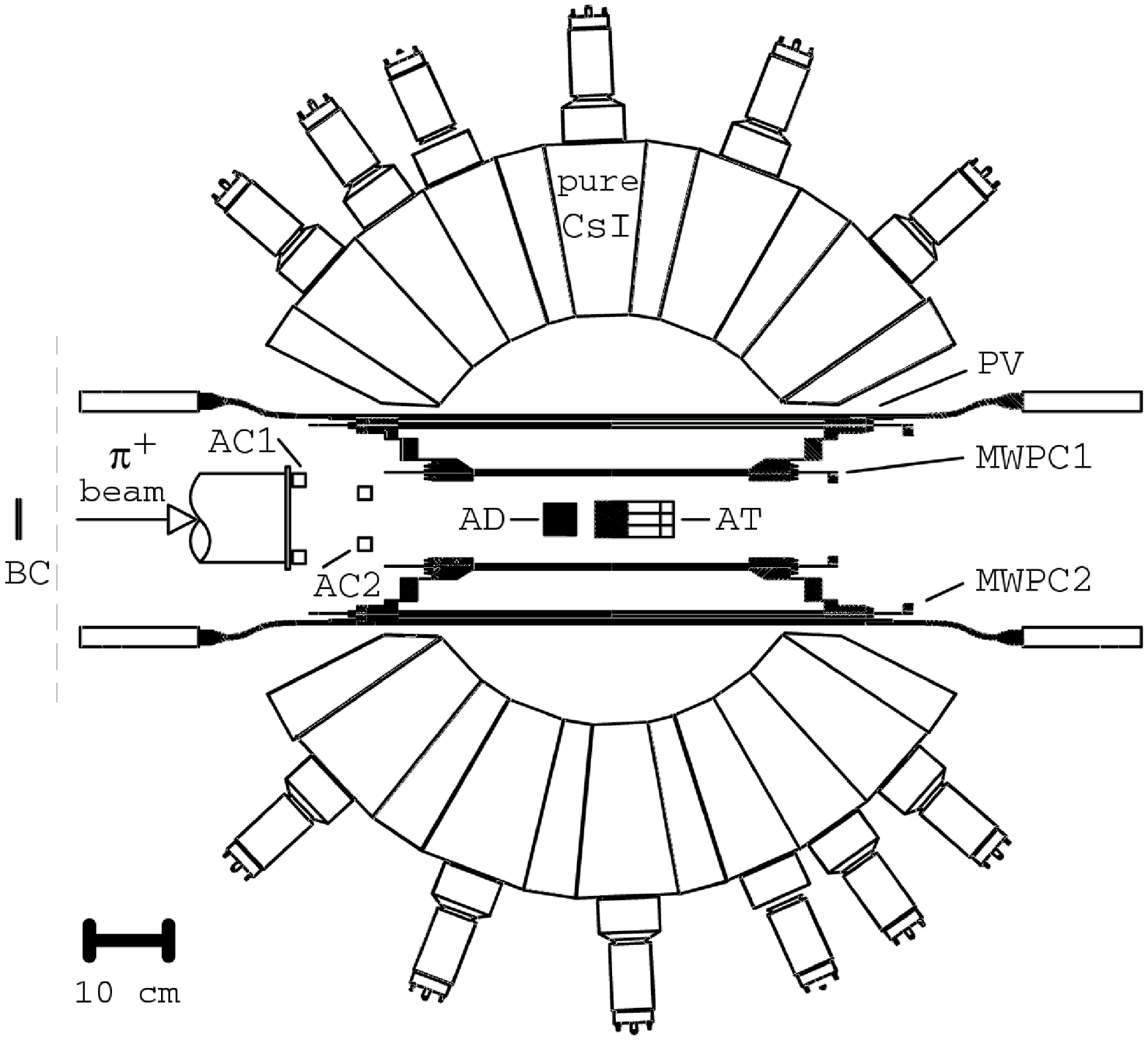}
\includegraphics[angle=0, height=.275\textheight]{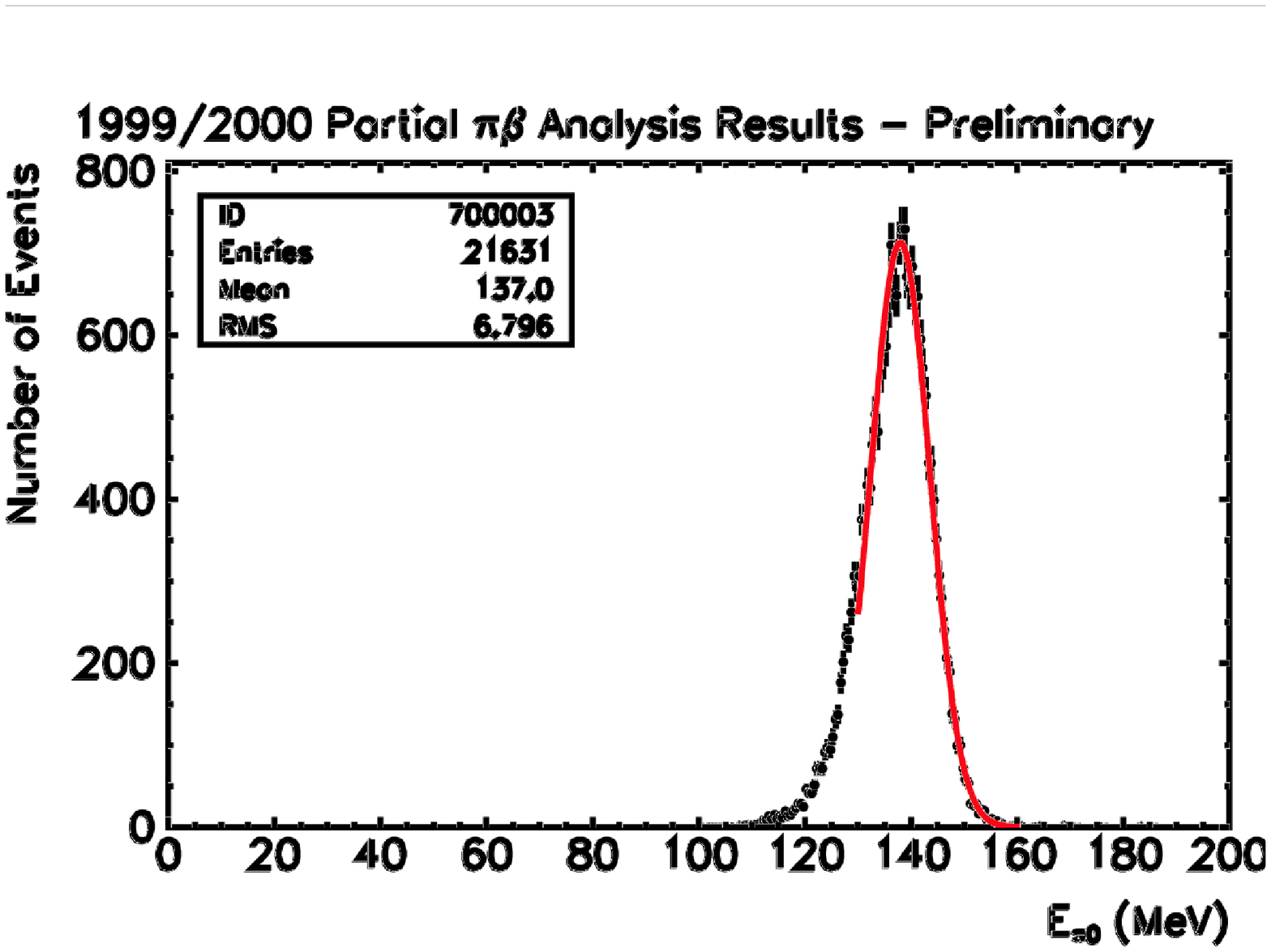}
  \caption{{\bf Left:} Layout of the PIBETA $\pi^+ \to \pi^0 e^+ \nu$ experiment.
{\bf Right:} $\pi^0$ energy spectrum for $\pi^+ \to \pi^0 e^+ \nu$ candidates from
PIBETA experiment.
    \label{fig:pibeta} }
\end{figure}

Fig.~\ref{fig:pibeta}-left is a layout of the PIBETA experiment.
A $\sim 1\,$MHz low energy $\pi^+$ beam is slowed from 113 MeV/c to 92
MeV/c in an active degrader and stopped in a 9-element segmented active 
target.  Decays at rest are detected during a
delayed $\sim 7 \tau_{\pi^+}$ gate.  Photons are detected in a
240-element CsI-pure array.  One is basically looking for $\pi^0$'s in
a small energy range following the decay of a $\pi^+$.  Detection of
the very soft positron is not required, although it is often observed
and used for systematic checks.  The detector includes a tracking
chamber surrounding the stopping target, that is used for other
physics quarries of the experiment, such as $\pi^+ \to e^+ \nu
(\gamma)$ and $\mu^+ \to e^+ \nu_e \bar \nu_{\mu}
\gamma$. Fig.~\ref{fig:pibeta}-right shows the $\pi^0$ energy spectrum
from the 1999/2000 run.  This is quite an impressive signal for a
$10^{-8}$-level decay.  The branching ratio is normalized via a
measurement of the decay $\pi^+ \to e^+ \nu$. There is a preliminary
result $B(\pi^+ \to \pi^0 e^+ \nu) = (1.044 \pm 0.007_{stat} \pm
0.015_{sys}) \times 10^{-8}$~\cite{Pocanic:2002}.  This is not yet
sufficient to influence the unitarity problem, but it represents a
factor 3 improvement on the previous
measurement\cite{Mcfarlane:1983eb}.  Further data is under analysis
and an eventual statistical sensitivity of 0.33\% is expected.  An
overall precision of $\leq\,$0.5\% is expected.  To go significantly
beyond this, one must reduce the systematic error due to the current
precision on the normalizing branching ratio of $\pi^+ \to e^+ \nu$,
which was last measured some ten years 
ago~\cite{Britton:1992pg,Britton:1994cj,Czapek:1993kc}.
A separate experiment to improve the precision on this mode which would
allow the full potential of PIBETA to be realized is in the planning 
stage.  This decay is very interesting in its own right, in that it
can severely constrain (or uncover) BSM physics
by probing the limits of lepton universality~\cite{Bryman:1993gm}.

\vspace{1.5cm}
\centerline{$K^+ \to \ell^+ \nu_{\ell} e{^+} e{^-}$}
\vspace{1.0cm}

Recently AGS E865 has published data on the decays $K^+ \to \mu^+
\nu_{\mu} e^+ e^-$ and $K^+ \to e^+ \nu_{e} e^+
e^-$~\cite{Poblaguev:2002ug}.  These decays can proceed via inner
bremsstrahlung (IB) off the $\ell^+$ or the $K^+$ in $K^+ \to \ell^+
\nu_{\ell}$ or, more interestingly from the point of view of
$\chi$PT, via structure dependent (SD)
radiation~\cite{Bijnens:1993en}.  There can also be a contributions 
from the interference of these two amplitudes.  
At leading order in $\chi$PT, the SD part is 0 so the
decays go entirely by IB.  At $\mathcal{O}(p^4)$, the SD contribution
is finite and can be characterized by constant form factors $F_V$,
$F_A$, $R$, where the latter is related to the kaon charge radius.  In
principle there can also be a tensor amplitude characterized by a form
factor $F_T$, although this is not allowed in the SM.  However there
have been hints of such an interaction in other semileptonic weak
decays~\cite{Bolotov:1990yq,Akimenko:1991fv}, so that it is of
interest to allow for it in analyzing $K^+ \to \ell^+ \nu_{\ell} e^+
e^-$.  At higher order in $\chi$PT, the form factors can be functions of 
$W^2$ and $q^2$, the effective mass squares of the $\ell^+
\nu_{\ell}$ system and the $e^+ e^-$ pair respectively.  The
IB term is helicity suppressed by a large factor in $K^+ \to e^+ \nu_e
e^+ e^-$ but dominates $K^+ \to \mu^+ \nu_{\mu} e^+ e^-$.  However one
is sensitive to all three terms (IB, SD and interference) in the
large $q^2$ region and so can hope to extract the signs of the form factors
relative to that of the kaon decay constant, $F_K$.

\begin{figure}[h]
\includegraphics[angle=0, height=.25\textheight]{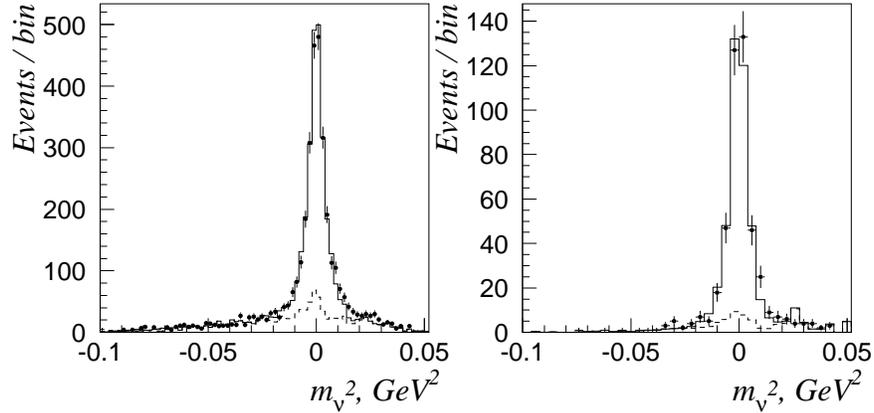}
  \caption{Missing mass distributions for {\bf left:} $K^+ \to \mu^+ \nu_{\mu} e^+ 
e^-$ and {\bf right:} $K^+ \to e^+ \nu_e e^+ e^-$.  Dashed lines indicate 
background and solid lines simulated data.
    \label{fig:klnee} }
\end{figure}

	Previous data on these modes have been limited to 4 events of
$K^+ \to e^+ \nu_e e^+ e^-$ and 14 events of $K^+ \to \mu^+ \nu_{\mu}
e^+ e^-$~\cite{Diamant-Berger:1976hk}. E865 observed 410 $K^+ \to e^+
\nu_e e^+ e^-$ candidates including an estimated background of 40
events and 2679 $K^+ \to \mu^+ \nu_{\mu} e^+ e^-$ candidates including
an estimated background of 514 events.  Fig.~\ref{fig:klnee} shows the
missing mass distributions for the two samples.  The corresponding
measured branching ratios were $(2.48 \pm 0.14_{stat} \pm 0.14_{syst})
\times 10^{-8}$ ($m_{ee}>150\,$MeV) and $(7.06 \pm 0.16_{stat} 
\pm 0.26_{syst}) \times 10^{-8}$ ($m_{ee}>145\,$MeV) 
respectively~\cite{Poblaguev:2002ug}. 

Table~\ref{klnee} shows the results of a combined likelihood analysis by E865
in which $K^+ \to e^+ \nu_e e^+ e^-$ and $K^+ \to \mu^+ \nu_{\mu} e^+ e^-$
were fit simultaneously.  This analysis  assumed a resonance
dominated form factor dependence on $q^2$ (the $\rho(770)$) and $W^2$
(the $K^*(892)$ for $F_V$ and the $K_1(1270)$ for $F_A$ and $R$).  
A fit assuming constant form factors had substantially lower likelihood.
The ``expected'' values listed in the table are those of
$\mathcal{O}(p^4)$ $\chi$PT supplemented by data from pion decay and
in the case of $R$, from the measured value of the kaon charge radius
~\cite{Amendolia:1986ui}.  The tensor form factor is consistent with 
0, although the precision is not sufficient to rule out the effect
postulated as an interpretation~\cite{Poblaguev:1990tv} of a result on 
$\pi^- \to e^- \bar\nu_e  \gamma$~\cite{Bolotov:1990yq}.  A new result
on $\pi^+ \to e^+ \nu_e  \gamma$ which will bear on this question is 
expected soon from the PIBETA experiment.

\begin{table}[h]
\begin{tabular}{cccc} \hline
form factor & value $\pm$ stat,sys,model & ``expected'' value & $\pi^+$ value \\
\hline
$F_V$ & $112 \pm 15 \pm 10 \pm 3$ & $96$  & $60 \pm 28$ \\
$F_A$ & $35 \pm 14 \pm 13 \pm 3$ & $41 \pm 6$  & $41 \pm 6$ \\
$R$ & $227 \pm 13 \pm 10 \pm 9$ & $230 \pm 34$  & $209 \pm 30$ \\
$F_T$ & $-4 \pm 7 \pm 7 \pm 0.4$ & $0$  & $-5.6 \pm 1.7$ \\
\end{tabular}
\caption{\it Results of a combined form factor analysis of $K^+ \to \ell^+ 
\nu_{\ell} e^+ e^-$ by E865 compared with expectations and with results from 
the pion sector. Units are $10^{-3}$.}
\label{klnee}
\end{table}

\section{Conclusions}

The success of lepton flavor violation experiments in reaching
sensitivities corresponding to mass scales of well over 100 TeV has
helped kill most models predicting accessible LFV in kaon decay.  Thus
new dedicated experiments in this area are unlikely in the near
future.  Since the most sensitive LFV limits in pion decay are
parasitic to kaon experiments, similar remarks apply to them.

        The existing precision measurement of \kmm\ will be very useful if
theorists can make enough progress on calculating the dispersive 
long-distance amplitude, perhaps helped by experimental progress in
$K_L \to \gamma \ell^+ \ell^-$, $K_L \to 4~leptons$, etc.  The exploitation
of \kmm~ would also be aided by higher precision measurements of
some of the normalizing reactions, such as $K_L \to \gamma\gamma$.

        \kpnnp\ will clearly be further exploited.  Two
coordinated initiatives are devoted to this: a $10^{-11}$/event experiment
(E949) just underway at the BNL AGS and a $10^{-12}$/event
experiment (CKM) recently approved for the FNAL Main Injector.  The first
dedicated experiment to seek \kpnn0~(E391a) is proceeding and an experiment (KOPIO)
at the AGS with the goal of making a $\sim 10\%$ measurement of
$Im(\lambda_t)$ is approved and in R\&D.

	Measurements of \kpnnp~and \kpnn0~can determine an
alternative unitarity triangle that will offer a critical comparison
with results from the $B$ system.  If new physics is in play in the
flavor sector, the two triangles will almost certainly disagree.

	$K_L \to \pi^0 \ell^+ \ell^-$ will probably not be further pursued 
unless and until a BSM signal is seen in $K \to \pi \nu\bar\nu$.

\begin{theacknowledgments}
I thank D. Bryman, M. Diwan, S. Kettell, W. Marciano, D. Pocanic, R. Shrock,
G. Valencia,
and M. Zeller for useful discussions, access to results, and other
materials. This work was supported by the U.S.  Department of Energy
under Contract No. DE-AC02-98CH10886.
\end{theacknowledgments}


\doingARLO[\bibliographystyle{aipproc}]
          {\ifthenelse{\equal{\AIPcitestyleselect}{num}}
             {\bibliographystyle{arlonum}}
             {\bibliographystyle{arlobib}}
          }
\bibliography{zuoz}

\end{document}